\documentclass[iop]{emulateapj}

\slugcomment{Accepted for publication in \textit{The Astrophysical Journal}}

\shorttitle{\textit{XMM} CLUSTER SURVEY: AGN AND STARBURSTS IN J$2215.9-1738$}
\shortauthors{HILTON ET AL.}

\begin{document}

\title{The \textit{XMM} Cluster Survey: Active Galactic Nuclei and Starburst Galaxies in XMMXCS
J2215.9-1738 at \lowercase{$z=1.46$}}

\author{Matt Hilton\altaffilmark{1}, Ed Lloyd-Davies\altaffilmark{2}, S. Adam Stanford\altaffilmark{3,4}, John
P. Stott\altaffilmark{5}, Chris A. Collins\altaffilmark{5}, A. Kathy Romer\altaffilmark{2},
Mark Hosmer\altaffilmark{2}, Ben Hoyle\altaffilmark{6}, Scott T. Kay\altaffilmark{7}, Andrew R.
Liddle\altaffilmark{2}, Nicola Mehrtens\altaffilmark{2}, Christopher J. Miller\altaffilmark{8}, Martin
Sahl{\'e}n\altaffilmark{9}, Pedro T. P. Viana\altaffilmark{10,11}}
\affil{
$^{1}$Astrophysics \& Cosmology Research Unit, School of Mathematical Sciences, University of KwaZulu-Natal,
Private Bag X54001, Durban 4000, South Africa; hiltonm@ukzn.ac.za\\
$^{2}$Astronomy Centre, University of Sussex, Falmer, Brighton, BN1 9QH, UK\\
$^{3}$University of California, Davis, CA 95616, USA\\
$^{4}$Institute of Geophysics and Planetary Physics, Lawrence Livermore National Laboratory, Livermore, CA
94551, USA\\
$^{5}$Astrophysics Research Institute, Liverpool John Moores University, Twelve Quays House, Egerton Wharf,
Birkenhead, CH41 1LD, UK\\
$^{6}$Institut de Ci\`{e}ncies del Cosmos (ICCUB), Departmento de F\'{\i}sica, Mart\'{\i} i Franqu\'{e}s 1,
08034 Barcelona, Spain\\
$^{7}$University of Manchester, Jodrell Bank Observatory, Macclesfield, Cheshire, SK11 9DL, UK\\
$^{8}$Astronomy Department, University of Michigan, Ann Arbor, MI 48109, USA\\
$^{9}$The Oskar Klein Centre for Cosmoparticle Physics, Department of Physics, Stockholm University, AlbaNova,
SE-106 91 Stockholm, Sweden\\
$^{10}$Departamento de F\`{\i}sica e Astronomia da Faculdade de Ci\^{e}ncias da Universidade do Porto, Rua do
Campo Alegre, 687, 4169-007, Portugal\\
$^{11}$Centro de Astrof\'{\i}sica da Universidade do Porto, Rua das Estrelas, 4150-762, Porto, Portugal\\
}

\begin{abstract}
We use \textit{Chandra} X-ray and \textit{Spitzer} infrared observations to explore the AGN and starburst
populations of XMMXCS J2215.9-1738 at $z=1.46$, one of the most distant spectroscopically confirmed galaxy
clusters known. The high resolution X-ray imaging reveals that the cluster emission is
contaminated by point sources that were not resolved in \textit{XMM-Newton} observations of the system, and
have the effect of hardening the spectrum, leading to the previously reported temperature for this system
being overestimated. From a joint spectroscopic analysis of the \textit{Chandra} and \textit{XMM-Newton} data,
the cluster is found to have temperature $T=4.1_{-0.9}^{+0.6}$~keV and luminosity $L_{\rm X} =
(2.92_{-0.35}^{+0.24}) \times 10^{44}$~erg~s$^{-1}$, extrapolated to a radius of 2~Mpc. As a result of this
revised analysis, the cluster is found to lie on the $\sigma_v-T$ relation, but the cluster remains less
luminous than would be expected from self-similar evolution of the local $L_{\rm X}-T$ relation. Two of the
newly discovered X-ray AGN are cluster members, while a third object, which is also a prominent 24~$\micron$
source, is found to have properties consistent with it being a high redshift, highly obscured object in the
background. We find a total of eight $>5\sigma$ 24~$\micron$ sources associated with cluster members (four
spectroscopically confirmed, and four selected using photometric redshifts), and one additional 24~$\micron$
source with two possible optical/near-infrared counterparts that may be associated with the cluster. Examining
the IRAC colors of these sources, we find one object is likely to be an AGN. Assuming that the other
24~$\micron$ sources are powered by star formation, their infrared luminosities imply star formation rates
$\sim100$~M$_\sun$~yr$^{-1}$. We find that three of these sources are located at projected distances of $<
250$~kpc from the cluster center, suggesting that a large amount of star formation may be taking place in the
cluster core, in contrast to clusters at low redshift.
\end{abstract}


\section{Introduction}
\label{s_intro}
The universe at high redshift is a much more active place than we see locally. Radio and far-infrared studies
have shown that the comoving star formation rate density increases by a factor $\approx 10$ from $z=0$ to $z
\approx 1$ \citep[e.g.][]{Seymour_2008, Magnelli_2009}, and continues rising towards higher redshifts, peaking
at $z \approx 3$ \citep{Bouwens_2009}. The fraction of galaxies with Active Galactic Nuclei (AGN) is also seen
to be much higher when the universe was young: current observations suggest that the space density of the
brightest AGNs peaks at $z \approx 2-3$ \citep{Assef_2010, Silverman_2005}. Observations in the local
universe show that the stellar mass of galactic bulges is correlated with the mass of nuclear
super-massive black holes \citep[e.g.][]{Ferrarese_2000, Gebhardt_2000}, suggesting that AGN play a major
role in the growth of stellar mass. Simulations suggest that interactions and mergers
between gas-rich galaxies drive this process, by providing both fuel for AGNs and triggering starburst
activity in galactic nuclei \citep[e.g.][]{Granato_2004, Hopkins_2008}. Eventually, gas heating by the AGN
(feedback) reaches such a level that star formation is shut down, leaving the stellar population of the host
galaxy to evolve passively. The quenching of star formation in massive galaxies by AGN feedback is a crucial
ingredient in the latest semi-analytic models of galaxy formation, bringing the models into much closer
agreement with observations \citep[e.g.][]{DeLucia_2006, Croton_2006, Bower_2006, Somerville_2008}.

A connection between local environment, star formation and AGN is therefore expected in this theoretical
picture. Intriguingly, recent studies of the dependence of star formation rate (SFR) on local galaxy density
using large samples of field galaxies indicate that SFR increases with increasing local density at $z \approx
1$ \citep{Elbaz_2007, Cooper_2008} -- whereas the opposite is observed in the local universe
\citep[e.g.][]{Lewis_2002, Balogh_2004}. This situation is mirrored to some extent in galaxy clusters.
Although these systems are dominated by quiescent, early type galaxies -- primarily
`red-and-dead' ellipticals -- and continue to host substantial populations of these objects up to the highest
redshifts observed \citep[e.g.][]{Blakeslee_2003, Lidman_2008, Mei_2009, Hilton_2009}, an increasing fraction
of star forming, late type galaxies are seen in clusters at higher redshifts \citep[e.g][]{ButcherOemler_1984,
vanDokkum_2000, Ellingson_2001, Smith_2005}. However, recent mid-infrared observations of clusters conducted
using the \textit{Spitzer Space Telescope} show that a significant amount of this increased star formation at
high redshift is obscured by dust \citep{Geach_2006, Marcillac_2007, Saintonge_2008, Haines_2009}, and
therefore the amount of star formation in clusters has previously been underestimated.

The AGN fraction in clusters is also seen to rise with redshift. \citet*{Martini_2009} report an eight-fold
increase in the number of AGNs in clusters at $z=1$ compared to in the local universe. Similarly,
\citet{Galametz_2009} find that IR-selected clusters in the IRAC Shallow Cluster Survey
\citep{Eisenhardt_2008} show a clear excess of AGNs at high redshift, with the AGN fraction
increasing by a factor of three from $0.5 < z < 1.0$ to $1.0 < z < 1.5$. However, it is not yet clear if the
AGN fraction is higher in clusters than in the field, nor if the AGN fraction is evolving more rapidly in
dense environments.

\defcitealias{Stanford_2006}{S06}
\defcitealias{Hilton_2007}{H07}
\defcitealias{Hilton_2009}{H09}

In this paper we use \textit{Chandra} X-ray observations and mid-infrared observations taken using the
\textit{Spitzer Space Telescope} to characterize the AGN and dusty star forming population of the $z=1.46$
cluster XMMXCS J2215.9-1738 (hereafter J2215), which was discovered in the ongoing optical follow-up of the
\textit{XMM} Cluster Survey \citep[XCS;][]{Romer_2001, Sahlen_2009}. J2215 is the second highest
redshift X-ray selected cluster known, following the recent discovery of a $z=1.62$ cluster in the
Subaru/\textit{XMM-Newton} Deep Field (\citealt*{Tanaka_2010}, \citealt{Papovich_2010}). The discovery of
J2215 and an analysis of its X-ray properties using \textit{XMM-Newton} data was reported in
\citet[][hereafter S06]{Stanford_2006}. We presented a first study of the dynamical state of the cluster and
measurement of its velocity dispersion using additional VLT and Keck spectroscopy in \citet[][hereafter
H07]{Hilton_2007}. Most recently, we have used \textit{Hubble Space Telescope} (\textit{HST}) imaging and
ground-based near-infrared data obtained at the \textit{Subaru} telescope to perform a detailed examination of
the morphologies of the cluster galaxies and the red-sequence within $<0.75$~Mpc of the cluster core
\citep[][hereafter H09]{Hilton_2009}.

The structure of this paper is as follows. In Section~\ref{s_XRayAnalysis}, we report revised
measurements of the cluster X-ray properties obtained through a joint spectroscopic analysis of the
\textit{Chandra} and \textit{XMM-Newton} data, taking into account the effect of X-ray point sources that were
unresolved in the \textit{XMM-Newton} data used in \citetalias{Stanford_2006}. In
Section~\ref{s_imaging} we briefly review the optical and ground based near-infrared data used in this
paper, before describing new photometry obtained using the Infrared Array Camera \citep[IRAC;][]{Fazio_2004}
that extends wavelength coverage of the cluster into the rest frame near-infrared. We describe additional
spectroscopic observations obtained at the Keck and Gemini observatories during 2008-2009
in Section~\ref{s_OpticalSpectroscopy}, where we also present an updated measurement of the cluster
velocity dispersion. Mid-infrared (24~$\micron$) observations of the cluster obtained using the Multiband
Imaging Photometer for \textit{Spitzer} (MIPS) are described in Section~\ref{s_MIPSAnalysis}. We
characterize the properties of cluster 24~$\micron$ and X-ray sources in terms of star formation or AGN
activity in Section~\ref{s_starFormation}. Finally, we discuss our findings in comparison with lower
redshift studies of clusters in Section~\ref{s_discussion}.

We assume a concordance cosmology of $\Omega_m=0.3$, $\Omega_\Lambda=0.7$, and $H_0=70$~km~s$^{-1}$~Mpc$^{-1}$
throughout, where $\Omega_\Lambda$ is the energy density associated with a cosmological constant. All
magnitudes are on the AB system \citep{Oke_1974}, unless otherwise stated.

\section{X-ray Observations}
\label{s_XRayAnalysis}

\begin{figure}
\begin{center}
\includegraphics[width=8.5cm]{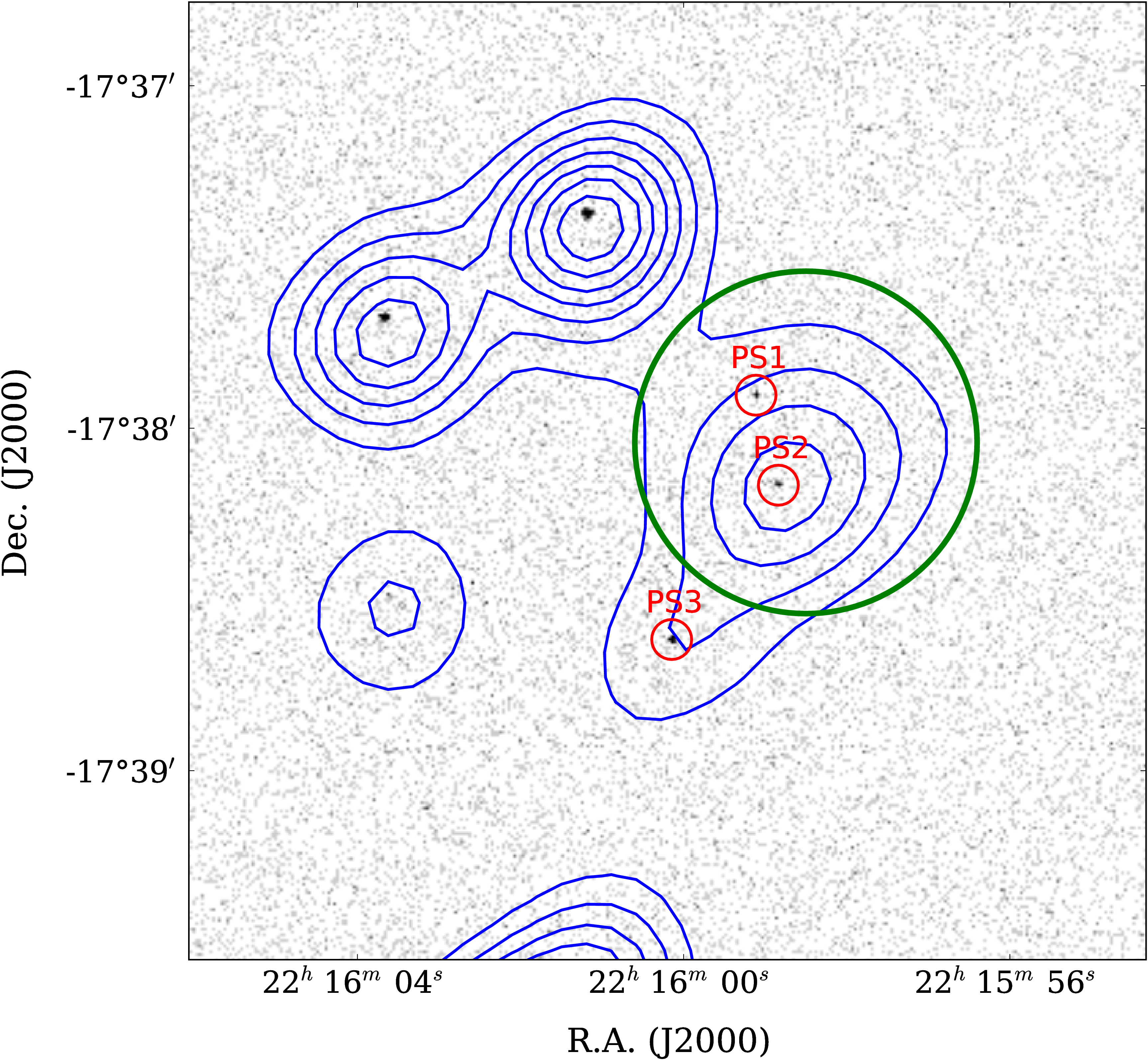}
\end{center}
\caption{\textit{Chandra} ACIS-S image of J2215.9-1738 with the \textit{XMM-Newton} EPIC contours
\citep{Stanford_2006} overlaid. The large circle marks the 30$\arcsec$ radius extraction region used
for the \textit{XMM-Newton} spectroscopic analysis in both \citet{Stanford_2006} and this work. The two
unlabeled point sources to the North East were previously known from the \textit{XMM-Newton} observations 
and taken into account in the analysis presented in \citet{Stanford_2006}. The three previously unknown 
point sources which contaminated the original \textit{XMM} detection are labelled PS1--3.
\label{fig1}}
\end{figure}

\subsection{Joint \textit{XMM-Chandra} Spectroscopic Analysis}
\label{s_jointXRay}
J2215 is detected as an extended X-ray source in the XCS, but the low resolution of \textit{XMM-Newton},
especially off-axis, could lead to significant contamination of the cluster emission by unresolved point
sources. The comoving space density of bright AGN rises steeply beyond $z > 0.5$ and at $z \approx 1.5$ is
nearing its peak \citep{Silverman_2005}. There is therefore a much higher probability of a high
redshift cluster such as J2215 containing an X-ray bright AGN compared to clusters in the local universe.

To check for point source contamination, we obtained a 56.2~ks \textit{Chandra} ACIS-S observation of the
cluster\footnotemark, resulting in 55.3~ks after cleaning to remove periods of high
background. The ACIS-S image reveals two point sources within the 30$\arcsec$ radius extraction region used
for the \textit{XMM-Newton} spectroscopic analysis described in \citetalias{Stanford_2006}. Another
previously unidentified point source is detected $\approx 12\arcsec$ outside the extraction region used in
the \citetalias{Stanford_2006} analysis, but some flux from this object is spread into the extraction region
by the larger PSF of \textit{XMM-Newton}.
\footnotetext{\textit{Chandra} ObsID 7919 and 8566}

Figure \ref{fig1} shows the ACIS-S image with the \textit{XMM-Newton} contours overlaid (newly
identified point sources in the \textit{Chandra} data are marked PS1-3). The three sources are detected with
only 23, 27 and 48 counts respectively, but despite this it is possible to fit their spectra with simple
power law models including photoelectric absorption (WABS; using the \texttt{XSPEC} package,
\citealt{Arnaud_1996}) in order to constrain their properties. The results of the spectral fitting are
shown in Table~\ref{tab1}.

The spectral properties are consistent with the sources being AGN, with power law indices in the
typical range for these objects. The fluxes of the sources span the range $(1.0-1.3) \times
10^{-15}$~erg~s$^{-1}$~cm$^{-2}$ in the (0.5--2.0)~keV band. Sources PS1 and PS2 are spectroscopically
confirmed cluster members which show 3727 \AA{} \textsc{[O ii]} emission but not broad lines (see
Section~\ref{s_OpticalSpectroscopy}). Source PS3 is also a $24~\micron$ source (see
Section~\ref{s_MIPSAnalysis}), but is not spectroscopically confirmed despite 8 hours of integration on this
object being obtained at Gemini (see Section~\ref{s_OpticalSpectroscopy}, below). The object does have a low
confidence photometric redshift from the analysis of \citetalias{Hilton_2009} placing the object in the
background of the cluster at $z_p \approx 2.5$. Assuming these redshifts, we estimate luminosities for the
X-ray point sources after removing the effects of intrinsic absorption within the host galaxies. We find
that the sources have unabsorbed luminosities spanning the range $L_{\rm X(2-10 \ keV)} \approx (0.1-6) \times
10^{44}$~erg~s$^{-1}$, typical of AGN rather than starburst galaxies. We list the source properties in
Table~\ref{tab1}, and discuss the properties of the AGN further in Section~\ref{s_AGN} below.

In order to calculate the fraction of flux from these point sources contaminating the cluster emission, we
used the \texttt{SCISIM}\footnotemark package to produce simulated images of the sources detected by
\textit{Chandra} blurred by the \textit{XMM-Newton} PSF. The estimated fraction of the flux from each point
source that falls within the \textit{XMM-Newton} extraction region is also listed in Table \ref{tab1}. We find
that the point sources contribute $\approx 15\%$ of the total flux measured within the extraction region.
\footnotetext{\url{http://xmm.vilspa.esa.es/scisim/release/}}

\begin{deluxetable*}{ccccccccccc}
\tablewidth{0pt}
\tabletypesize{\scriptsize}
\tablecaption{Properties of X-ray point sources from \textit{Chandra} ACIS-S spectroscopy\label{tab1}}
\tablehead{\colhead{ID} &  \colhead{R.A.} & \colhead{Dec.} & \colhead{$z$} & \colhead{$N_{\rm
H}$\tablenotemark{a}} &
\colhead{$\alpha$} & \colhead{$f_{\rm X}$\tablenotemark{b}} & \colhead{EF\tablenotemark{c}} &
\colhead{CF\tablenotemark{d}} & \colhead{$L_{\rm X(0.5-2.0 \ keV)}$\tablenotemark{e}} & \colhead{$L_{\rm
X(2-10 \ keV)}$\tablenotemark{e}}}
\startdata
PS1 &    22 15 59.115 & -17 37 54.24 & 1.462 & 1.6$^{+2.6}_{-0.8}$  & 2.3$_{-0.8}^{+1.1}$ & 1.0 & 0.75 &
0.065 & 0.36 & 0.25\\
PS2 &    22 15 58.857 & -17 38 09.98 & 1.453 & 0.67$^{+0.91}_{-0.47}$    & 4.1$_{-0.8}^{+1.2}$ & 1.3 & 
0.63 & 0.071 & 1.2 & 0.14\\
PS3 &    22 16 00.148 & -17 38 37.04 & 2.5\tablenotemark{*}  & 44$^{+27}_{-14}$ & 1.9$_{-0.7}^{+0.8}$ &
1.3 & 0.09 & 0.009 & 5.4 & 6.7\\
\enddata
\tablecomments{Units of right ascension are hours, minutes and seconds, and units of declination are 
degrees, minutes, and seconds (J2000).}
\tablenotetext{a}{Hydrogen column density ($\times 10^{22}$~cm$^{-2}$).}
\tablenotetext{b}{Flux ($\times 10^{-15}$~erg~s$^{-1}$~cm$^{-2}$), in the (0.5-2.0)~keV band. The 
uncertainty in these values is $\approx$30--50\%.}
\tablenotetext{c}{The fraction of the source flux that is estimated to be within the extraction region used
for the \textit{XMM-Newton} spectroscopic analysis.}
\tablenotetext{d}{Estimate of the fraction that the source contributes to the cluster flux measured by
\citet{Stanford_2006}.}
\tablenotetext{e}{Unabsorbed luminosity ($\times 10^{44}$~erg~s$^{-1}$) in the indicated energy band.}
\tablenotetext{*}{The redshift for PS3 is a low confidence photometric estimate.}
\end{deluxetable*}

\begin{figure}
\begin{center}
\includegraphics[width=8cm]{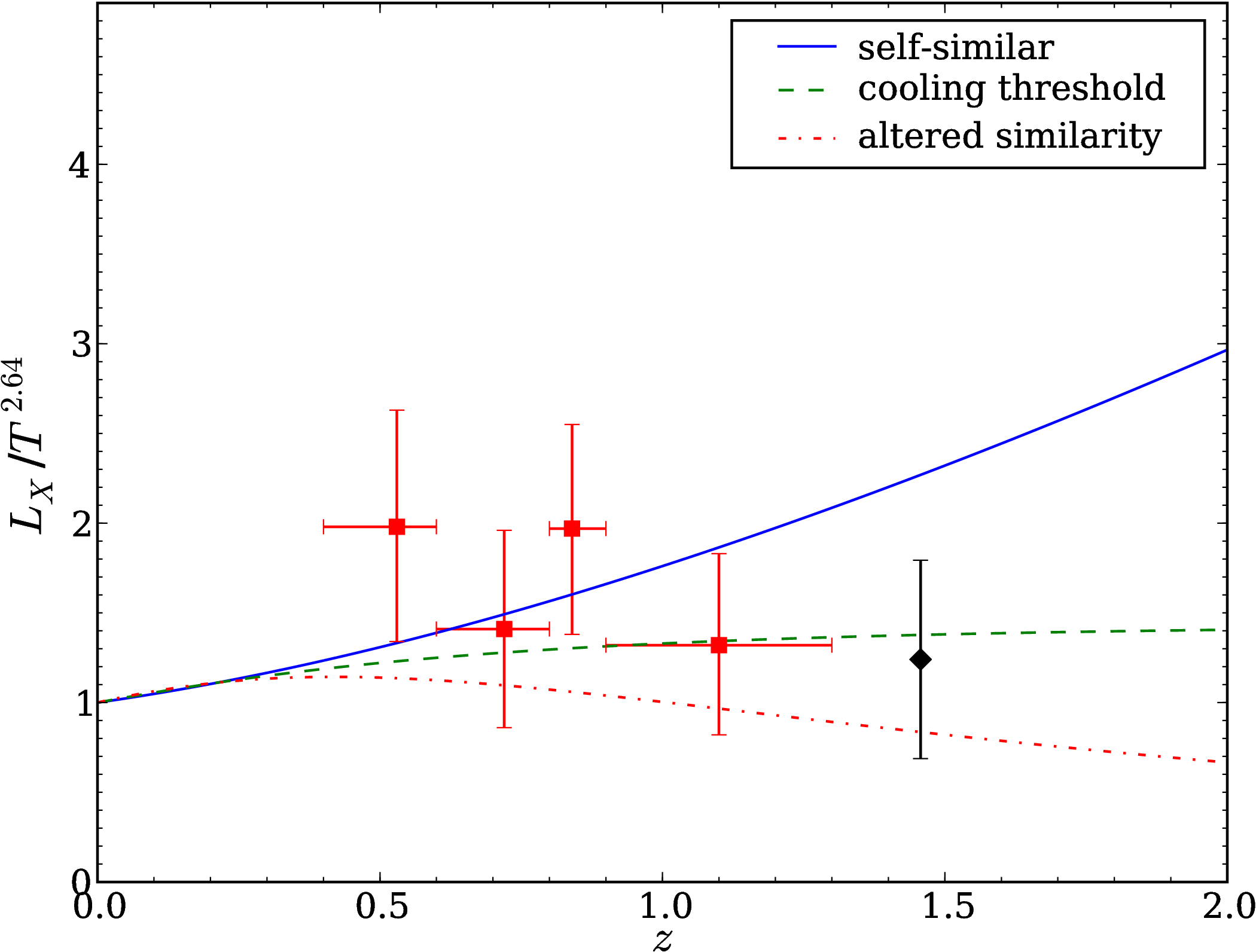}
\end{center}
\caption{Comparison of $L_{\rm X}/T^{2.64}$ (i.e., assuming the slope of the \citealt{Markevitch_1998} 
$L_{\rm X}-T$ relation) for J2215.9-1738 (diamond point; the error bar is estimated using a bootstrap
resampling technique) with the predicted evolution of the normalization of the $L_{\rm X}-T$ relation 
for the cases of self-similarity ($E(z)$), cooling threshold ($t_0 / \left[ E(z) t(z) \right]$), and 
altered similarity ($t_0^2 / \left[ E(z)^3 t(z)^2 \right]$; see \citealt{Voit_2005} for a description of
these latter two models). The square points are the data of \citet{Maughan_2006}; vertical error bars 
are equal to the weighted standard deviation at each redshift, horizontal error bars indicate the width 
of each redshift bin.}
\label{f_LTNorm}
\end{figure}

To take into account the effect of the point source contamination on the cluster properties measured using
\textit{XMM-Newton} \citepalias{Stanford_2006}, it was necessary to undertake a joint fit of the
\textit{XMM-Newton} and \textit{Chandra} data. A model consisting of three absorbed powerlaw plus an absorbed
MEKAL (Mewe-Kaastra-Liedahl) plasma model was jointly fitted to the \textit{XMM-Newton} spectra from the
extraction region and \textit{Chandra} spectra of the individual point sources with the flux fractions in the
extraction region calculated from \texttt{SCISIM} taken into account. The resulting temperature from this
analysis is $T=4.1_{-0.9}^{+0.6}$~keV.

A temperature $T=7.4^{+2.7}_{-1.8}$~keV (90\% confidence) was measured in the analysis
of \citetalias{Stanford_2006} based on the \textit{XMM-Newton} data alone. \citetalias{Stanford_2006} also
considered the possibility of contamination from an unresolved central point source, estimating a lower
temperature of $T=6.5^{+2.6}_{-1.8}$~keV (90\% confidence) in that case. Both of these temperature estimates
are somewhat higher than the temperature measured from the joint \textit{Chandra}-\textit{XMM} analysis
presented here.

The bolometric luminosity of the cluster is found to be $L_{\rm X} = (2.92_{-0.35}^{+0.24}) \times
10^{44}$~erg~s$^{-1}$, extrapolated to a radius of 2~Mpc assuming a $\beta$ model with $\beta=0.63$ and core
radius $r_c = 52$~kpc. This is $\approx 33$\% lower than the equivalent measurement reported in
\citetalias{Stanford_2006}. As discussed in \citetalias{Hilton_2007}, a comparison of the
\citetalias{Stanford_2006} luminosity and temperature measurements indicates that the cluster is underluminous
compared to the value expected if the $L_{\rm X}-T$ relation evolves self-similarly with redshift. The cluster
was found to follow the evolution expected in the \citet{Voit_2005} analytic models that include the effects
of non-gravitational heating and radiative cooling \citep[see also][and the discussion in
Hilton et al. 2007]{VoitPonman_2003}. We find that this remains the case using the luminosity and temperature
estimates obtained in the joint \textit{Chandra}-\textit{XMM} analysis presented here, although the size of
the deviation from the self-similar $E(z)$ scaling is reduced to the $\approx 2\sigma$ level. This is
illustrated in Figure~\ref{f_LTNorm}, which shows the predicted evolution of the normalization of the $L_{\rm
X}-T$ relation for the self-similar case, in comparison to the `cooling threshold' ($t_0 / \left[ E(z)
t(z) \right]$) and `altered similarity' ($t_0^2 / \left[ E(z)^3 t(z)^2 \right]$) models of \citet{Voit_2005}.
We assume the slope of the $L_{\rm X}-T$ relation measured by \citet{Markevitch_1998} in comparing the
data to the models. The uncertainty on the value of $L_{\rm X}/T^{2.64}$ inferred from the J2215
luminosity and temperature measurements is estimated using a bootstrap resampling technique.

\begin{figure}[b]
\begin{center}
\includegraphics[width=8cm]{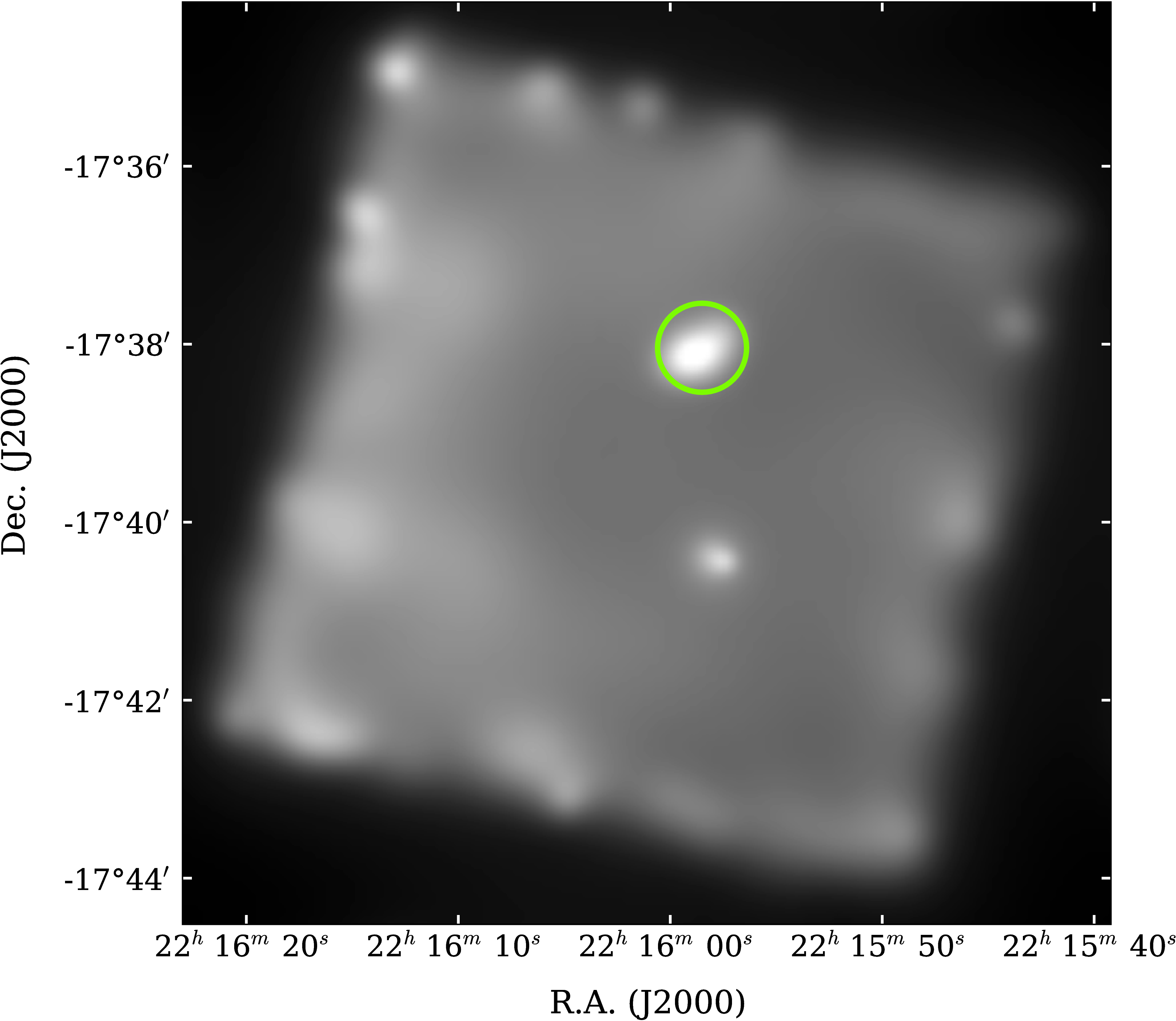}
\end{center}
\caption{Adaptively smoothed (after point source subtraction) \textit{Chandra} ACIS-S image of the field 
of J2215.9-1738. The 30$\arcsec$ radius extraction region used for the \textit{XMM-Newton} spectroscopic
analysis is shown by the circle. The cluster extended emission is clearly visible, and is detected at the
$\approx 9\sigma$ level. Another faint extended source, BLOX J2215.9-1740.4 \citep[][which is also 
detected in the XCS extended source catalog]{Dietrich_2007} is clearly visible below J2215.9-1738. 
No other significant sources are seen.}
\label{fig2}
\end{figure}

\subsection{Extended Emission}
In addition to studying the point sources in the region of J2215, for which the \textit{Chandra} telescope is
most suited due to its high spatial resolution, we also attempted to identify extended emission from the
cluster in the \textit{Chandra} imaging data alone. Given the considerably lower effective area of
\textit{Chandra} and significantly lower exposure time for these observations compared to the
\textit{XMM-Newton} observations, the amount of information that can obtained from such a detection is
limited, but does provide an independent confirmation of the cluster X-ray emission. In order to study
extended emission in the \textit{Chandra} data we first removed flux due to the detected point sources (an
easy task given the small, well characterized point spread function of \textit{Chandra}), and produced a
(0.3-2.0~keV) band image (the energy band was chosen to maximize the signal-to-noise). The \texttt{CIAO} task
\texttt{csmooth} was then used to adaptively smooth the image.

The resulting smoothed image can be seen in Figure \ref{fig2}, where we have highlighted the 30$\arcsec$
radius spectral extraction region used for the \textit{XMM-Newton} spectroscopic analysis. It can be
seen that there is a strong extended source coincident with the position of J2215. The extended emission
is detected at the $\approx 9\sigma$ level, with 123 counts in the (0.3-2.0 keV) band. Unfortunately, due
to the low number of counts, we are unable to place comparable constraints on the cluster temperature to
those obtained with the \textit{XMM-Newton} data from the \textit{Chandra} data alone, although the spectrum
is consistent with the temperature being in the 3-6~keV range.

\section{Optical and Near-Infrared Imaging}
\label{s_imaging}

We have previously obtained imaging of J2215 at both optical and near-infrared wavelengths using a variety of
different telescopes and instruments. Wide field coverage is provided by $I$-band data taken from the ESO
Imaging Survey \citep{Dietrich_2006}, and by $K_s$-band imaging from the Palomar 5~m telescope: these data are
described in \citetalias{Stanford_2006}, and the overlapping area between them covers an $8\arcmin \times
8\arcmin$ area centered on the cluster. The $K_s$-band data in this case reach a 5$\sigma$ limiting magnitude
of $\approx 21.8$ (AB). Much deeper observations of the central $3\arcmin \times 3\arcmin$ of the cluster
field have been obtained using \textit{HST} ($i_{775}$, $z_{850}$) and Subaru ($J$, $K_s$). These data reach
$\approx 2.5$ magnitudes deeper at $K_s$ than the Palomar data, and were used to construct a photometric
redshift catalog \citepalias{Hilton_2009}. The \textit{HST} data were obtained as part of a supernova search
in high-redshift clusters \citep{Dawson_2009}. In the present study we add imaging with the \textit{Spitzer}
IRAC instrument to this dataset, which extends the wavelength coverage into the rest-frame near-infrared at
the cluster redshift.

\subsection{Spitzer IRAC Data}
\label{s_IRAC}
IRAC imaging of J2215 was obtained through Program 50333 using all four channels (3.6, 4.5, 5.8,
8.0~$\micron$) on 2008 July 12. The frame time was 100~s and 15 exposures were obtained in a medium scale
cycling dither pattern. The Basic Calibrated Data (BCD) images were corrected for muxbleed and pulldown and
then mosaiced using \texttt{MOPEX} \citep{MakovozKhan_2005}.

Photometry was performed using \texttt{SExtractor} \citep{BertinArnouts_1996} in a two stage process. In the
first stage, an initial set of catalogs were generated for each IRAC band, which were then merged by
associating objects located within $<1.2\arcsec$ of each other across the catalogs (note the astrometry of
the IRAC mosaics is accurate to $<0.1\arcsec$). The positions of objects in these catalogs were
used to register the IRAC mosaics to a common pixel coordinate system. These registered mosaics were then
trimmed to a region of $5.9\arcmin \times 6.6\arcmin$ with coverage in all IRAC bands, centered on the cluster
X-ray position. In the second stage, we ran \texttt{SExtractor} in dual image mode, extracting a 3.6~$\micron$
selected catalog from the registered and trimmed IRAC images in each band using a consistent set of
photometric apertures.

We measured fluxes through $4\arcsec$ diameter circular apertures, adopting aperture corrections to total
magnitudes of $-0.350\pm 0.040$, $-0.350\pm0.040$, $-0.540 \pm0.040$, $-0.660 \pm 0.020$ magnitudes for IRAC
channels 1--4 respectively, as measured by \citet[][note that aperture corrections consistent
with these values were also found by \citealt{Papovich_2006} and \citealt{PerezGonzalez_2008}]{Barmby_2008}.
We use these aperture corrected magnitudes to measure colors between the IRAC bands (see
Section~\ref{s_AGN}). 

The uncertainties in the aperture corrections were added in quadrature to the photometric uncertainties
reported by \texttt{SExtractor}; however, at this stage the photometric errors are underestimated due to the
effect of resampling the IRAC images during the production of the mosaics, which introduces correlations in
the noise between adjacent pixels. This additional source of uncertainty is not taken into account in the
\texttt{SExtractor} photometric errors. We correct for this effect using simulations, in a similar manner to
that described by \citet{Barmby_2008}. We measure the background flux through 100 non-overlapping $4\arcsec$
diameter apertures placed at random locations in each IRAC mosaic that do not coincide with detected objects,
repeating this process 1000 times. We scale the aperture corrected \texttt{SExtractor} photometric errors by
the median rms background flux measured from these simulations divided by the median \texttt{SExtractor} flux
error of the detected objects. The scale factors applied to the uncertainties in the total aperture
magnitudes were 2.2, 1.8, 2.5, 1.9 across IRAC channels 1--4 respectively.

The IRAC catalog was cross matched against the $K_s$-selected catalogs described above using a 1.2$\arcsec$
matching radius. Clearly, given the much larger PSF of IRAC ($1.67-1.98\arcsec$) compared to these other
catalogs, many sources are blended in the cluster center: in each case we choose the nearest match, and will
note such affected objects where necessary in the subsequent analysis.

\section{Optical Spectroscopy}
\label{s_OpticalSpectroscopy}
\subsection{Observations and Data Reduction}
Redshifts for 17 galaxies concordant with the cluster redshift were reported in \citetalias{Hilton_2007}. We
have observed J2215 using a further six slit masks since that work: five with the DEep Imaging and
Multi-Object Spectrograph \citep[DEIMOS;][]{Faber_2003} on the 10~m Keck II telescope, plus one additional
mask observed with the Gemini Multi-Object Spectrograph \citep[GMOS;][]{Hook_2003} on the 8~m Gemini South
telescope. Redshifts for galaxies within $3\arcmin$ of the cluster center from three of the additional DEIMOS
masks (obtained in 2007 September and 2008 September) were included in the catalog included in
\citetalias{Hilton_2009}. The two DEIMOS masks and the GMOS mask observed in 2009 were designed to prioritize
objects detected in our mid-infrared imaging (see Section~\ref{s_MIPSAnalysis} below). We now describe
the observations and data reduction for each instrument in turn.

\subsubsection{DEIMOS}
\label{s_DEIMOS}
For all masks, DEIMOS was used with the 600ZD grating, which is blazed at 7500~\AA{} and has dispersion
0.65~\AA{}~pixel$^{-1}$, in conjunction with the OG515 order sorting filter. This setup provides typical
wavelength coverage from $5000-10000$~\AA{}. Slits of width $1.2\arcsec$ and minimum length $5\arcsec$ were
used in all slit masks. Each mask contained typically $\approx 110$ slitlets, spread across the $16.7\arcmin
\times 5.0\arcmin$ field of view of DEIMOS.

A total of five masks were observed with DEIMOS subsequent to the analysis presented in
\citetalias{Hilton_2007}. We observed one mask on UT 2007 September 10, obtaining $3 \times 1800$ sec
exposures in clear conditions with $0.7\arcsec$ seeing. Two masks were observed on UT 2008 September 01-02 in
mostly clear conditions. The seeing was typically $0.6\arcsec$ on September 01, rising to $0.9\arcsec$ on the
following night. In these masks we preferentially targeted galaxies close to the red-sequence with photometric
redshifts concordant with the cluster redshift from the catalog presented in \citetalias{Hilton_2009}. We
obtained $5 \times 1800$ sec of integration time through the mask observed on September 01 and $7 \times 1800$
sec through the second mask, which was observed on the following night. Finally, on UT 2009 October 18-19 we
observed two more masks, in which 24~$\micron$ sources (see Section~\ref{s_MIPSAnalysis} below) were
prioritized. These observations were carried out in clear conditions with $0.7-0.8\arcsec$ seeing. We obtained
$5 \times 1800$ sec exposures for each of these masks, observing one mask on each night.

The DEIMOS data were reduced as described in \citetalias{Hilton_2007}, using version 1.1.4 of \texttt{spec2d},
the pipeline developed for the DEEP2 galaxy redshift survey \citep{Davis_2003}.

\subsubsection{GMOS}

\begin{deluxetable*}{cccccccccccc}
\tablewidth{0pt}
\tabletypesize{\scriptsize}
\tablecaption{Galaxies within $\pm 3000$~km~s$^{-1}$ of the recession velocity corresponding to the redshift
of J2215.9-1738 ($z=1.46$)\label{t_redshifts}}
\tablehead{\colhead{ID}& \colhead{R.A. (J2000)}& \colhead{Dec. (J2000)}& \colhead{$z$\tablenotemark{a}}&
\colhead{Q\tablenotemark{b}} & \colhead{Method\tablenotemark{c}} & \colhead{$I$\tablenotemark{d}} &
\colhead{$K_s$\tablenotemark{e}} &
\colhead{H09\tablenotemark{f}} & \colhead{r (arcmin)} & \colhead{r (Mpc)} & \colhead{[O
\textsc{ii}]\tablenotemark{g}}}
\startdata
1 & 22 15 58.464 & -17 37 58.58 & 1.452 & 3 & V & 23.63 & 20.90 & 742 & 0.066 & 0.033 & \nodata\\
3 & 22 15 58.872 & -17 37 59.26 & 1.451 & 3 & V &23.30 & 21.05 & 748 & 0.104 & 0.053 & \nodata\\
4 & 22 15 58.056 & -17 38 04.59 & 1.465 & 2 & V &\nodata & 23.60 & 789 & 0.111 & 0.057 & \nodata\\
5 & 22 15 59.040 & -17 38 02.54 & 1.454 & 2 & V &22.72 & 20.86 & 770 & 0.129 & 0.065 & $\checkmark$\\
6 & 22 15 58.488 & -17 38 10.53 & 1.465 & 3 & X &23.73 & 20.57 & 824 & 0.134 & 0.068 & \nodata\\
7 & 22 15 59.080 & -17 38 02.40 & 1.459 & 2 & V &22.72 & 20.86 & 770 & 0.138 & 0.070 & $\checkmark$\\
10 & 22 15 58.848 & -17 38 09.85 & 1.453 & 2 & V &22.75 & 21.77 & 821 & 0.148 & 0.075 & $\checkmark$\\
16 & 22 15 59.136 & -17 37 54.22 & 1.462 & 3 & X &24.36 & 20.72 & 712 & 0.205 & 0.104 & $\checkmark$\\
18 & 22 15 59.376 & -17 38 09.88 & 1.467 & 3 & V &24.56 & 23.89 & 823 & 0.242 & 0.123 & $\checkmark$\\
19 & 22 15 57.432 & -17 37 57.90 & 1.454 & 2 & V &\nodata & 23.05 & 754 & 0.266 & 0.135 & $\checkmark$\\
30 & 22 15 57.192 & -17 38 07.80 & 1.450 & 3 & X &23.24 & 22.05 & 810 & 0.324 & 0.164 & $\checkmark$\\
35 & 22 15 57.240 & -17 37 53.22 & 1.454 & 3 & V &23.69 & 21.41 & 708 & 0.338 & 0.171 & $\checkmark$\\
36 & 22 15 57.720 & -17 37 45.55 & 1.451 & 2 & V &24.17 & 20.91 & 653 & 0.338 & 0.172 & $\checkmark$\\
38 & 22 15 58.752 & -17 37 40.76 & 1.460 & 2 & X &\nodata & 22.33 & 631 & 0.367 & 0.186 & $\checkmark$\\
39 & 22 15 56.928 & -17 38 04.70 & 1.445 & 3 & V &23.03 & 21.82 & 786 & 0.376 & 0.191 & $\checkmark$\\
44 & 22 15 58.344 & -17 37 37.30 & 1.453 & 3 & V &24.11 & 22.71 & 629 & 0.422 & 0.214 & $\checkmark$\\
53 & 22 16 00.384 & -17 37 50.59 & 1.451 & 3 & X &23.52 & 20.99 & 692 & 0.491 & 0.249 & $\checkmark$\\
57 & 22 16 00.648 & -17 37 51.67 & 1.471 & 2 & V &23.66 & 21.17 & 696 & 0.543 & 0.275 & \nodata\\
59 & 22 15 56.184 & -17 37 49.90 & 1.454 & 3 & X &23.64 & 20.57 & 688 & 0.590 & 0.300 & \nodata\\
62 & 22 15 56.016 & -17 37 55.05 & 1.460 & 2 & V &24.41 & 23.88 & 723 & 0.605 & 0.307 & \nodata\\
63 & 22 16 00.096 & -17 37 33.92 & 1.454 & 3 & V &\nodata & 22.86 & 591 & 0.609 & 0.309 & $\checkmark$\\
65 & 22 15 56.058 & -17 37 49.90 & 1.461 & 2 & V &\nodata & \nodata & \nodata & 0.618 & 0.314 & $\checkmark$\\
66 & 22 15 59.472 & -17 38 37.39 & 1.457 & 3 & X &23.09 & 22.77 & 1018 & 0.626 & 0.318 & $\checkmark$\\
77 & 22 16 00.096 & -17 38 42.61 & 1.463 & 2 & V &23.57 & 22.25 & 1047 & 0.769 & 0.390 & \nodata\\
82 & 22 15 56.544 & -17 37 21.36 & 1.446 & 2 & V &23.67 & 21.49 & 526 & 0.829 & 0.421 & $\checkmark$\\
105 & 22 16 00.360 & -17 38 57.69 & 1.456 & 2 & V &\nodata & 21.64 & 1130 & 1.021 & 0.518 & $\checkmark$\\
106 & 22 15 55.200 & -17 37 22.36 & 1.460 & 2 & V &23.65 & 21.36 & 538 & 1.032 & 0.524 & \nodata\\
120 & 22 16 03.432 & -17 37 51.24 & 1.462 & 2 & V &24.11 & 22.82 & 697 & 1.190 & 0.604 & $\checkmark$\\
161 & 22 15 54.000 & -17 39 01.65 & 1.464 & 2 & V &23.19 & 21.29 & 1149 & 1.457 & 0.739 & $\checkmark$\\
162 & 22 15 52.464 & -17 37 46.05 & 1.461 & 2 & V &22.77 & 20.99 & \nodata & 1.464 & 0.743 & $\checkmark$\\
176 & 22 15 56.160 & -17 36 37.80 & 1.455 & 3 & V &24.98 & 23.03 & 96 & 1.518 & 0.770 & $\checkmark$\\
203 & 22 16 05.400 & -17 38 17.48 & 1.465 & 3 & X &\nodata & 22.75 & 859 & 1.663 & 0.844 & $\checkmark$\\
232 & 22 15 51.648 & -17 37 11.74 & 1.461 & 3 & X &24.07 & 20.77 & \nodata & 1.839 & 0.933 & \nodata\\
264 & 22 16 02.448 & -17 39 53.46 & 1.456 & 2 & V &\nodata & 21.12 & \nodata & 2.075 & 1.053 & $\checkmark$\\
292 & 22 15 57.960 & -17 40 20.35 & 1.465 & 3 & V &22.17 & 19.36 & \nodata & 2.301 & 1.168 & \nodata\\
301 & 22 16 02.424 & -17 40 09.08 & 1.475 & 2 & X &22.67 & \nodata & \nodata & 2.308 & 1.171 & \nodata\\
303 & 22 15 54.864 & -17 40 12.97 & 1.475 & 2 & X &23.71 & \nodata & \nodata & 2.341 & 1.188 & \nodata\\
358 & 22 16 09.504 & -17 38 26.16 & 1.461 & 3 & X &23.81 & \nodata & \nodata & 2.651 & 1.345 & \nodata\\
371 & 22 16 09.888 & -17 38 26.41 & 1.459 & 2 & V &\nodata & 20.67 & \nodata & 2.742 & 1.392 & $\checkmark$\\
400 & 22 15 58.944 & -17 41 01.28 & 1.470 & 2 & V &23.41 & 21.33 & \nodata & 2.982 & 1.513 & $\checkmark$\\
439 & 22 16 10.296 & -17 36 22.96 & 1.463 & 3 & X &23.54 & \nodata & \nodata & 3.264 & 1.656 & $\checkmark$\\
494 & 22 16 10.728 & -17 40 24.31 & 1.463 & 3 & V &23.43 & 22.42 & \nodata & 3.751 & 1.904 & \nodata\\
505 & 22 16 04.104 & -17 34 23.37 & 1.474 & 3 & X &23.41 & \nodata & \nodata & 3.889 & 1.973 & $\checkmark$\\
883 & 22 15 50.832 & -17 29 09.45 & 1.458 & 3 & X &22.54 & \nodata & \nodata & 9.070 & 4.603 & $\checkmark$\\
\enddata
\tablecomments{Units of right ascension are hours, minutes and seconds, and units of declination are 
degrees, minutes, and seconds.}
\tablenotetext{a}{The redshift uncertainity is measured to be $1.4 \times 10^{-3}$, see
Section~\ref{s_redshifts}.}
\tablenotetext{b}{An additional 14 objects with $Q=1$ (insecure) redshifts are not listed in this table.}
\tablenotetext{c}{Redshift measurement method: V = from visually identified features; X = from cross
correlation with spectral templates (see Section~\ref{s_redshifts}).}
\tablenotetext{d}{Measured in ESO Imaging Survey data \citep{Dietrich_2006}; magnitudes have been converted 
to the AB system using $I(AB) = I(Vega) + 0.41$. If no value is given, the object was not detected. ID 65 is
detected in HST $i_{775}$ imaging of the cluster (described in \citealt{Hilton_2009}).}
\tablenotetext{e}{Taken from the catalog presented in \citet{Hilton_2009} where H09 ID is given; otherwise
measured from the Palomar $K_s$-band data described in \citet{Stanford_2006}. If no value is given, the 
object was not detected in either $K_s$-selected catalog.}
\tablenotetext{f}{ID in the catalog of \citet{Hilton_2009}.}
\tablenotetext{g}{$\checkmark$ indicates 3727 \AA{} [O \textsc{ii}] emission was identified in the object
spectrum.}
\end{deluxetable*}

Service mode observations of J2215 were obtained using GMOS over the period UT 2009 August 23--September 23.
GMOS was used in nod-and-shuffle mode \citep{GlazebrookBlandHawthorn_2001}, which allows both improved sky
subtraction and shorter slit lengths compared to more conventional techniques. We used the R400 grating, which
for our observations has dispersion 1.34~\AA{} pixel$^{-1}$ due to $2 \times 2$ binning of the detectors. The
OG515 order blocking filter was used, giving a nominal wavelength coverage of $5400-9700$~\AA{} when used in
combination with our chosen grating. One mask was observed, containing 34 target slitlets of length $3\arcsec$
and width $1\arcsec$. A total of $15 \times 1920$ sec exposures were obtained, divided equally among three
different central wavelengths (7500, 7550, 7600~\AA{}), in order to obtain coverage over the gaps between the
GMOS CCDs. We also varied the detector translation assembly offset by $\pm 2$ pixels in the $y$-direction to
mitigate against defects in the CCDs that cause artifacts in very deep nod-and-shuffle observations. The
nod-and-shuffle cycle time of our observations was 120 sec, and our offset size was chosen such that the
object was visible in the slit at both nod positions, to maximize the on-source integration time. Each of our
1920 sec exposures therefore consisted of 16 nod-and-shuffle cycles. The relevant observing constraints chosen
for the queue mode observations were: 85\%-ile image quality (i.e. seeing $< 1.1\arcsec$ in the $I$-band); and
50\%-ile sky transparency (i.e. photometric conditions).

The data were reduced using routines from v1.8 of the Gemini \texttt{IRAF}\footnotemark package. The data for
each night and grating angle combination were processed separately. Briefly, the data were bias subtracted in
the standard manner, sky subtracted (using the \texttt{gnsskysub} task), and then flatfielded. Data from
individual chips were then mosaiced together, before being cut into separate FITS extensions corresponding to
each slitlet using a custom written tool. Wavelength calibration was performed using a cross correlation
technique, which was found to be accurate to 0.3~\AA{}. All data were then combined using a median average,
rejecting bad pixels using a mask constructed from nod-and-shuffle dark frames. Finally, the pairs of spectra
corresponding to each nod position were combined appropriately and one-dimensional spectra were extracted
using a simple boxcar algorithm.
\footnotetext{\texttt{IRAF} is distributed by the National Optical Astronomy Observatories, which are operated
by the Association of Universities for Research in Astronomy, Inc., under cooperative agreement with the
National Science Foundation.}

\subsection{Redshift Measurements}
\label{s_redshifts}

Redshifts were measured using a similar technique to that described in \citetalias{Hilton_2007}. Briefly, the
spectra were cross correlated against a subset of the SDSS spectral templates\footnotemark, plus the Luminous
Red Galaxy template of \citet{Eisenstein_2003}, using the \texttt{xcsao} task in the \texttt{rvsao IRAF}
package \citep{KurtzMink_1998}. All spectra were visually inspected to check the accuracy of the redshift
solution; where the cross correlation redshift solution was clearly incorrect, the redshift was measured from
visually identified spectral features. Due to the limited wavelength coverage of the SDSS templates at blue
wavelengths in the rest frame at the cluster redshift, and the additional complication of residual sky
emission at wavelengths approaching $\sim 10000$~\AA{}, most ($\approx 2/3$) of the redshifts for cluster
members were measured by the visual inspection method rather than using the cross correlation method. However,
the cross correlation technique is very efficient for measuring the redshifts of lower redshift galaxies, of
which a significant number are contained in our masks, each of which typically targeted $\sim 100$ galaxies
out to $\sim 11\arcmin$ away from the cluster center.
\footnotetext{http://www.sdss.org/dr5/algorithms/spectemplates/index.html}

Redshifts were assigned a quality flag according to the following scheme: $Q=3$ (completely unambiguous, at
least two positively identified spectral features); $Q=2$ (high confidence that the redshift is correct, one
clearly detected feature); $Q=1$ (significant doubt that the redshift is correctly identified, one or
more weakly detected features). As discussed below, a significant fraction of the cluster members are
identified as 3727~\AA{} \textsc{[O~ii]} emitters, including 15 objects with $Q=2$ rather than $Q=3$
redshifts. Although the redshifts of these objects are in most cases determined from a single emission line,
we are reasonably confident that the line identification is correct, as no emission lines are seen blueward
down to $\sim 6000$~\AA{}.

A total of 16 objects now have multiple reasonably secure redshift measurements ($Q \geq 2$) across different
slit masks. We adopt the rms scatter of the residuals between these multiple redshift measurements as our
estimate of the total (i.e. random+systematic) redshift error, finding $\Delta z = 1.4 \times 10^{-3}$. We
list the redshifts of all galaxies found within $\pm 3000$~km~s$^{-1}$ of the recession velocity corresponding
to the cluster redshift of $z=1.46$ in Table~\ref{t_redshifts}. A total of 44 galaxies now have
$Q \geq 2$ redshifts in the range $1.435 < z < 1.485$, an increase of 26 galaxies over
the catalog presented in \citetalias{Hilton_2007}.

Approximately two thirds of the galaxies for which $Q \geq 2$ spectroscopic redshifts have been obtained show
3727 \AA{} \textsc{[O~ii]} emission (see Table~\ref{t_redshifts}). This includes 13 galaxies located within a
projected distance of $<250$~kpc from the cluster center, and implies that a significant amount of
star formation could be taking place even in the inner regions of the cluster, which is not found to be the
case in lower redshift clusters. However, several studies have shown that \textsc{[O~ii]} emission is often
not due to star formation at high redshift, and is instead associated with Seyfert-type AGNs or Low Ionization
Nuclear Emission Line Regions \citep[LINERs;][]{Heckman_1980}, which can also be found in otherwise passively
evolving galaxies \citep{Yan_2006, Lemaux_2010}. A significant fraction of \textsc{[O~ii]} emitting galaxies
in the cluster core has also been found by \citet{Hayashi_2010}, using narrow band imaging that reaches to an
\textsc{[O~ii]} flux limit $\geq 1.4 \times 10^{-17}$~erg~s$^{-1}$ (corresponding to equivalent width $\geq
35$~\AA{}), allowing galaxies with SFR $> 2.6$~M$_\sun$~yr$^{-1}$ to be detected. \citet{Hayashi_2010} report
that $\approx 30$\% of the cluster members within a projected distance of 250~kpc from the cluster center are
\textsc{[O~ii]} emitters. Given the heterogeneous nature of our spectroscopic target selection, and the
potential for selection biases (for example, the relative ease of identifying \textsc{[O~ii]} emission at the
cluster redshift compared to securing redshifts for passively evolving member galaxies), it is unfortunately
not possible to infer the fraction of star forming galaxies within the cluster based on our spectroscopic data
alone.

\subsection{Cluster Line of Sight Velocity Distribution}
\label{s_velDist}

\begin{figure}
\begin{center}
\includegraphics[width=8.5cm]{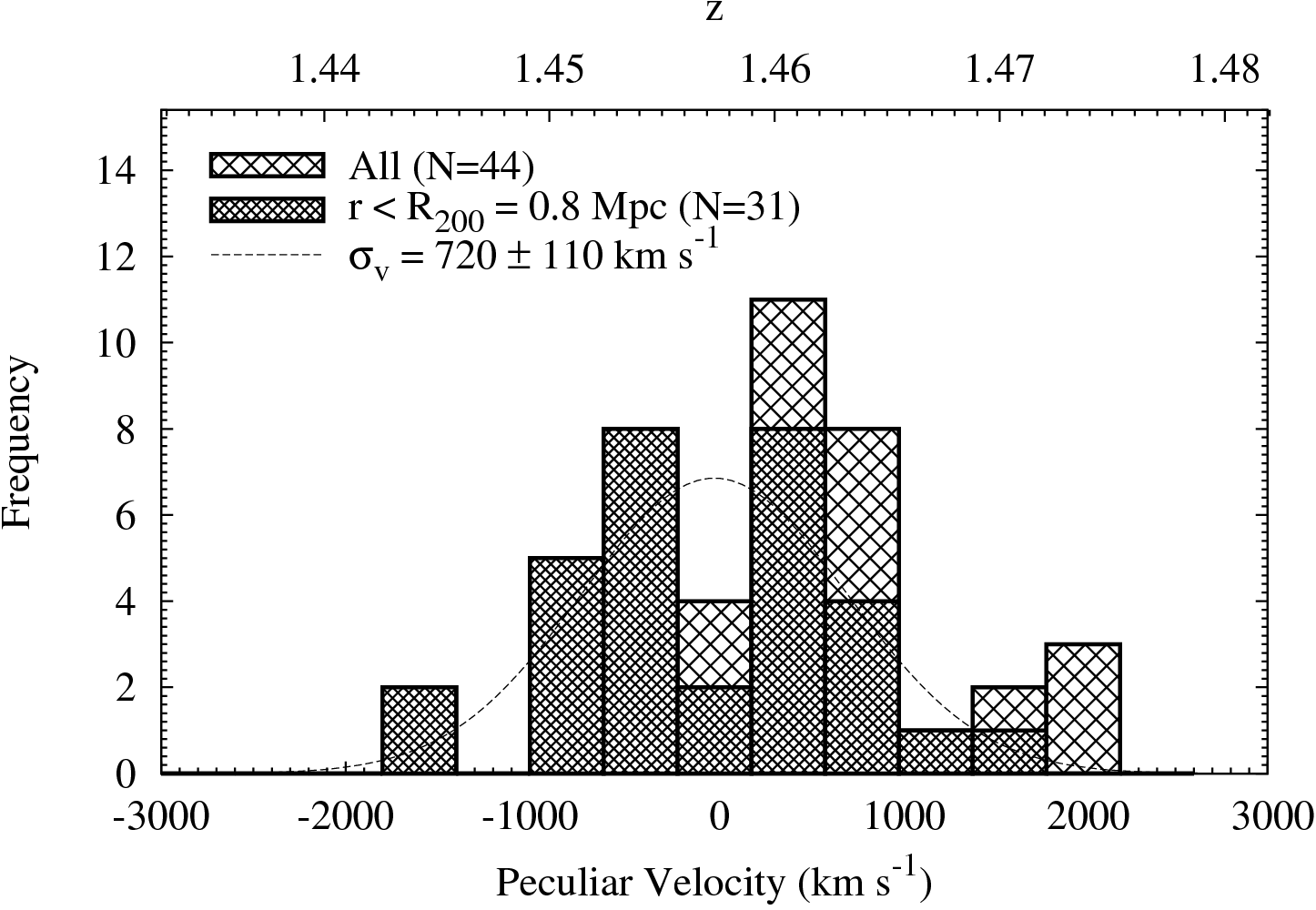}
\end{center}
\caption{The peculiar velocity distribution of J2215.9-1738. Only galaxies with reasonably secure ($Q \geq
2$) redshifts are included. The shading distinguishes between the sample of 31 galaxies located at radial
distance $r < R_{200}$ (see Section~\ref{s_velDist}), and the complete sample of 44 galaxies with 
$1.435 < z < 1.485$. The dashed line shows a Gaussian with $\sigma$ equal to the velocity dispersion
estimated using galaxies within $R_{200}$.}
\label{f_velDist}
\end{figure}

We plot the velocity distribution of all 44 galaxies with $Q \geq 2$ redshifts listed in
Table~\ref{t_redshifts} in Figure~\ref{f_velDist}. Calculating the velocity dispersion for the whole sample of
44 galaxies using a biweight scale estimator \citep{Beers_1990}, and applying the \citet*{Danese_1980}
correction for broadening of the distribution due to redshift measurement errors (see
Section~\ref{s_redshifts} above), we measure the line of sight velocity dispersion to be $\sigma_v = 890
\pm 110$ km~s$^{-1}$. However, this includes galaxies located up to $\approx 4.6$~Mpc away from the cluster
center, and not all galaxies at large radial distances may be gravitationally bound to the cluster. Using the
iterative method of calculating $\sigma_v$ and $R_{200}$ from the redshift data as described in
\citetalias{Hilton_2007}, we obtain a lower value of $\sigma_v = 720 \pm 110$ km~s$^{-1}$ using only the 31
galaxies within $R_{200}$, which we estimate to be $R_{200} = 0.8 \pm 0.1$ Mpc. We estimate the cluster
redshift to be $z=1.457 \pm 0.001$ by applying the biweight location estimator to this galaxy sample, in
agreement with the result of \citetalias{Hilton_2007}. We highlight the velocity distribution corresponding to
this membership selection in Figure~\ref{f_velDist}.

The value of $\sigma_v$ quoted here is higher than the value of $580 \pm 140$ km~s~$^{-1}$ estimated using the
smaller redshift catalog presented in \citetalias{Hilton_2007}, but agrees with the previous measurement at
better than $\lesssim 0.8\sigma$. In \citetalias{Hilton_2007}, we noted a discrepancy between the temperature
implied by the galaxy velocity dispersion and the X-ray temperature measured in \citetalias{Stanford_2006},
suggesting that the cluster gas has 2--3 times the kinetic energy of the cluster galaxies. The joint
\textit{XMM-Chandra} spectroscopic X-ray analysis presented in Section~\ref{s_XRayAnalysis} now
reconciles the velocity and X-ray measurements: for $\sigma_v = 720$ km~s$^{-1}$, the predicted X-ray
temperature is $T_{\sigma_v} = 3.2 \pm 0.9$~keV, compared to the revised X-ray measurement of
$T=4.1_{-0.9}^{+0.6}$~keV, and therefore agrees at $\lesssim 0.7 \sigma$. The cluster lies on the
$\sigma_v-T$ relation expected if there is equipartition of gravitational potential energy between the gas
and the galaxies following this revised analysis.

In \citetalias{Hilton_2007}, we also presented mild evidence that the cluster velocity distribution is
bimodal, and it is clear from inspection of Figure~\ref{f_velDist} that the additional redshift data has not
erased this signal. We performed the \citet{Hartigan2_1985} test of unimodality using the sample of 31
galaxies within $R_{200}$ as estimated above, measuring the value of the dip statistic to be 0.0668.
Simulations show that the probability to exceed this dip value when drawing from a Gaussian distribution with
$\sigma$ equal to the measured line of sight velocity dispersion is $P=0.103$. Therefore, the expanded
spectroscopic catalog does not strongly favor a bimodal velocity distribution using this sample. We also used
the \texttt{KMM} algorithm \citep*[Kaye's Mixture Model;][]{Ashman_1994} to test for bimodality. This
algorithm tests whether a mixture of $n$ Gaussian distributions is a better description of a given population
than a single Gaussian. For $n=2$, we found that the sample of 31 galaxies within $R_{200}$ does not favor a
bimodal Gaussian distribution ($P=0.603$), and this remains the case when considering galaxy
samples out to larger radius.

We conclude that despite the significantly increased size of the redshift catalog, there is no clear evidence
from the velocity distribution that the cluster consists of two kinematically distinct structures. However,
the fact that the \citet{Hartigan2_1985} dip test still mildly favors a bimodal distribution suggests that the
cluster may not be completely relaxed.

\section{Mid-IR Observations}
\label{s_MIPSAnalysis}

\subsection{Data Reduction and Photometry}
J2215 was observed at 24~$\micron$ using MIPS on 2008 June 21 as part of the same program that
obtained the IRAC observations described in Section~\ref{s_IRAC}. The angular resolution of MIPS is
$6\arcsec$, which corresponds to a physical scale of 51~kpc at the cluster redshift in our adopted
cosmology. A total of 60 frames each of 30 sec integration time were obtained using a large field size dither
pattern of 15 cycles centered on the cluster. The basic calibrated data (BCDs) as processed by the
\textit{Spitzer} Science Center were corrected for the effects of bright and dark latents, gradients, and the
``jailbar'' effect using a flatfield calibration generated from the data itself. The calibrated data were then
mosaiced together using the \texttt{MOPEX} package \citep{MakovozKhan_2005} to produce a final $5.5\arcmin
\times 6.3\arcmin$ image at 1.25$\arcsec$ pixelization roughly centered on the cluster position. A significant
fraction of the pixels in the top left corner of many of the individual BCD frames were flagged as suffering
from soft saturation, amounting to $\approx 14$\% of the total data collected. These pixels were rejected in
the mosaicing process and are therefore not included in the final science frame. Consequently, the noise level
is higher in this region, which is located $>2\arcmin$ away from the cluster position towards the North East.
As a result, the median exposure time in the final mosaic is reduced to $\approx 1620$ sec per pixel, although
the exposure time over $\approx 50$\% of the mosaic, including the cluster center, is $\approx 1800$ sec per
pixel.

Photometry was performed using the \texttt{APEX} package \citep{MakovozMarleau_2005}, where we adopt Point
Response Function (PRF) fitted fluxes as measurements of the total source flux, since at the cluster redshift
the member galaxies are unresolved by MIPS. The PRF model used was estimated from point sources detected at
$S/N > 10$ in an initial pass over the mosaic. We checked that the PRF estimate was of sufficient accuracy by
visual inspection of a residual image in which all detected point sources had been subtracted. Our final
24~$\micron$ catalog contains a total of 152 sources detected at $>5\sigma$.

We checked the flux detection and completeness limits of our 24~$\micron$ mosaic using simulations. We
inserted 10 artificial point sources, covering a wide range in flux, at random positions in the point source
subtracted mosaic (which should have the same noise properties as our science mosaic), and attempted to
recover them using the same extraction procedure used to make the catalog from the science mosaic. This
process was repeated 300 times. We estimate that the 50\% completeness limit of our observations is
approximately 70 $\mu$Jy on the basis of these simulations. Using the \citet{DaleHelou_2002} spectral
templates, this translates into a corresponding infrared luminosity limit of $L_{\rm IR} > 3 \times 10^{11}$
$L_\Sun$ at the cluster redshift. Assuming the \citet{Kennicutt_1998} law, this implies that we will detect
cluster members with SFR $> 60$ M$_\Sun$ yr$^{-1}$ (see Section~\ref{s_starFormation} below) at this
completeness level. Thus, we are only able to detect starburst galaxies at the cluster redshift.

\subsection{Source Matching}
\label{s_sourceMatching}
We cross matched the 24~$\micron$ catalog against the optical and near-IR catalogs described in
Section~\ref{s_imaging}, using a 2$\arcsec$ matching radius \citep{Bai_2007}. For some objects, multiple
possible counterparts were identified within the matching radius; we adopt the nearest source as the match
where this occurs. Matches were found for a total of 118 $>5\sigma$ 24~$\micron$ sources in this process.
We estimated the probability of spurious matches by randomizing the positions of all the 24~$\micron$
sources in the catalog and repeating the cross matching procedure. We estimate a $15\pm3$\% chance of a given
24~$\micron$ source being randomly associated with an optical/near-IR source, following 100 trials.

\begin{figure}
\begin{center}
\includegraphics[width=7cm]{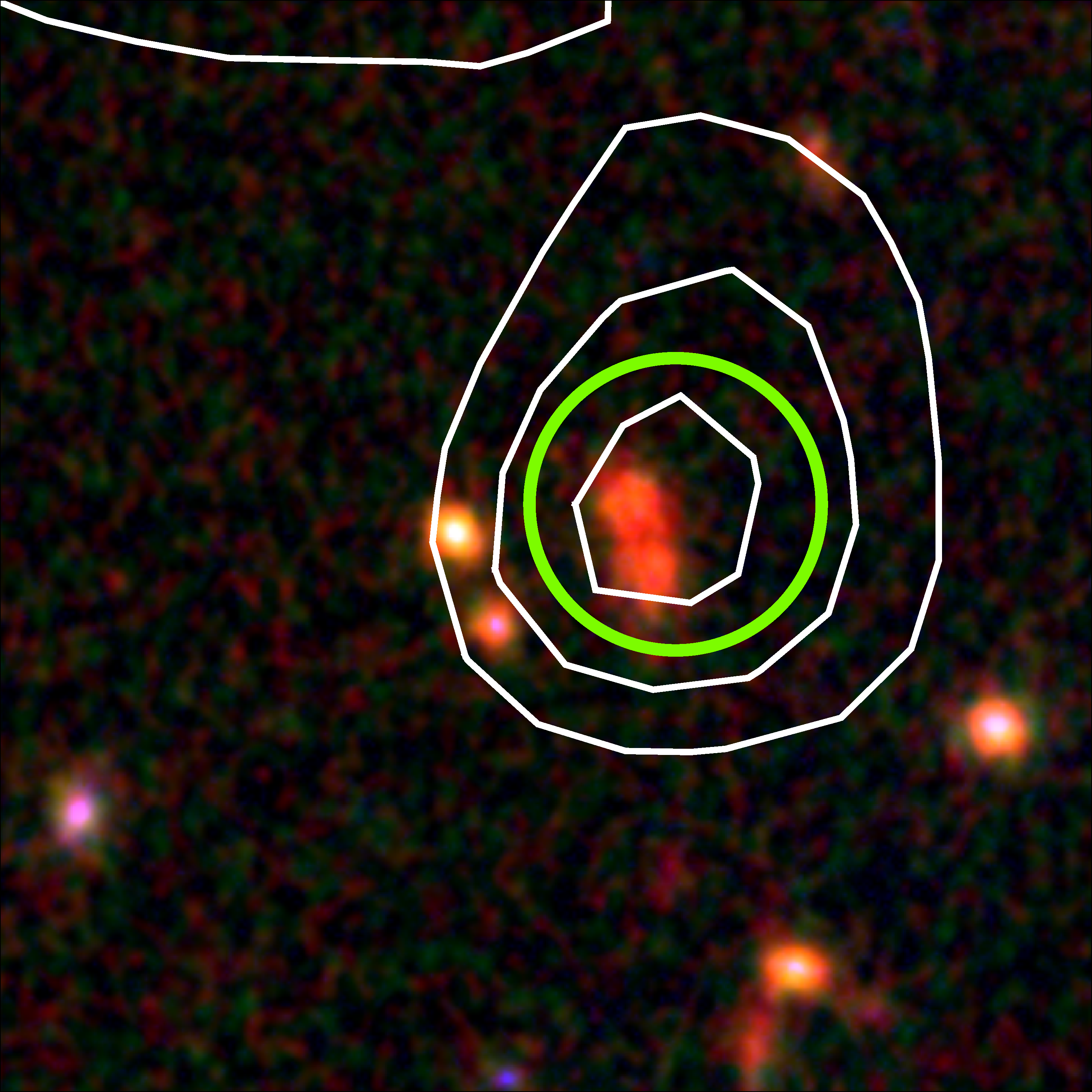}
\end{center}
\caption{False color image ($z_{850}$, $J$, $K_s$) of a zoomed in $15\arcsec \times 15 \arcsec$ region in 
the vicinity of a prominent, bright (239$\pm$9~$\mu$Jy) 24~$\micron$ source located $\approx 17\arcsec$ from
the cluster center (SSTU~J221559.68-173758.9). North is towards the top of the image, East is towards the
left. The contour overlay shows the MIPS 24~$\micron$ data. The green circle is centered on the position of
the 24~$\micron$ source and shows the $2\arcsec$ radius used in cross matching the 24~$\micron$ catalog to 
the optical/near-infrared catalogs (see Section~\ref{s_sourceMatching}). Although the source is not
spectroscopically confirmed as a cluster member, the two galaxies within the matching circle have a $Q=1$
(insecure) spectroscopic redshift and a photometric redshift concordant with the cluster redshift. In
addition, the two galaxies to the left of the source, just outside the matching radius, were selected as
members in the photometric redshift catalog of \citet{Hilton_2009}, suggesting that the source is likely to 
be associated with the cluster.}
\label{f_confused}
\end{figure}

Our interest is restricted to sources identified as cluster members, and we find that a total of four objects
in the spectroscopic redshift catalog (Table~\ref{t_redshifts}) are associated with $>5\sigma$ 24~$\micron$
detections. All of these objects show some evidence of \textsc{[O ii]} emission in their spectra. We
supplement this sample with a further four objects identified as being cluster members through their
photometric redshifts \citep[see][we assume these objects are all at $z=1.46$]{Hilton_2009}. The typical
scatter in the photometric redshift residuals ($\delta z = (z_s-z_p)/(1+z_s)$, where $z_s$
and $z_p$ are the spectroscopic and photometric redshifts respectively) is $\approx 0.04$ \citep{Hilton_2009}.
Note that one object thought to be a spurious detection has been removed from this sample after visual
inspection of all cluster members. None of the objects in this sample have alternative optical/near-infrared
counterparts within the 2$\arcsec$ matching radius used.

In addition to these objects, we found that there is a very prominent, bright (239$\pm$9~$\mu$Jy)
24~$\micron$ source located $17\arcsec$ away ($\approx 0.15$~Mpc at $z=1.46$) from the cluster center that is
not included by the above sample selection criteria. It has an insecure ($Q=1$)
spectroscopic redshift that hints it could be a cluster member, and in addition, is not unambiguously
identified with a single optical/near-infrared object: an object with a photometric redshift consistent with
cluster membership from the \citetalias{Hilton_2009} catalog is also located within the matching radius.
Furthermore, the local environment of this object is somewhat crowded, as Figure~\ref{f_confused} shows; there
are two other galaxies within $\lesssim 2.5\arcsec$ that were selected as photometric members of the cluster
in the \citetalias{Hilton_2009} catalog. This suggests that the source is likely to be physically associated
with the cluster. We will refer to this object (SSTU~J221559.68-173758.9) as the `unconfirmed source' in the
following discussion of the nature of the sources. It is possible that both of the candidate
optical/near-infrared counterparts of this object contribute to the measured 24~$\micron$ flux.

We also cross matched the MIPS source list against the X-ray point sources identified in our \textit{Chandra}
observations (Section~\ref{s_XRayAnalysis}, above). We find that none of the 24~$\micron$ sources
associated with cluster members are X-ray sources. However, one bright 24~$\micron$ source with flux density
$f_{\rm 24} = 820\pm 10$~$\mu$Jy (SSTU~J221600.14-173836.8) is associated with the \textit{Chandra} source
PS3 (see Table~\ref{tab1}), which is located only $\approx 0.7\arcmin$ from the cluster center. As mentioned
in Section~\ref{s_jointXRay}, this object is likely to be a high redshift ($z_p \sim 2.5$) obscured AGN
behind the cluster.

\section{Nature of the Cluster X-ray and Mid-infrared Sources}
\label{s_starFormation}

\begin{figure}
\begin{center}
\includegraphics[width=8cm]{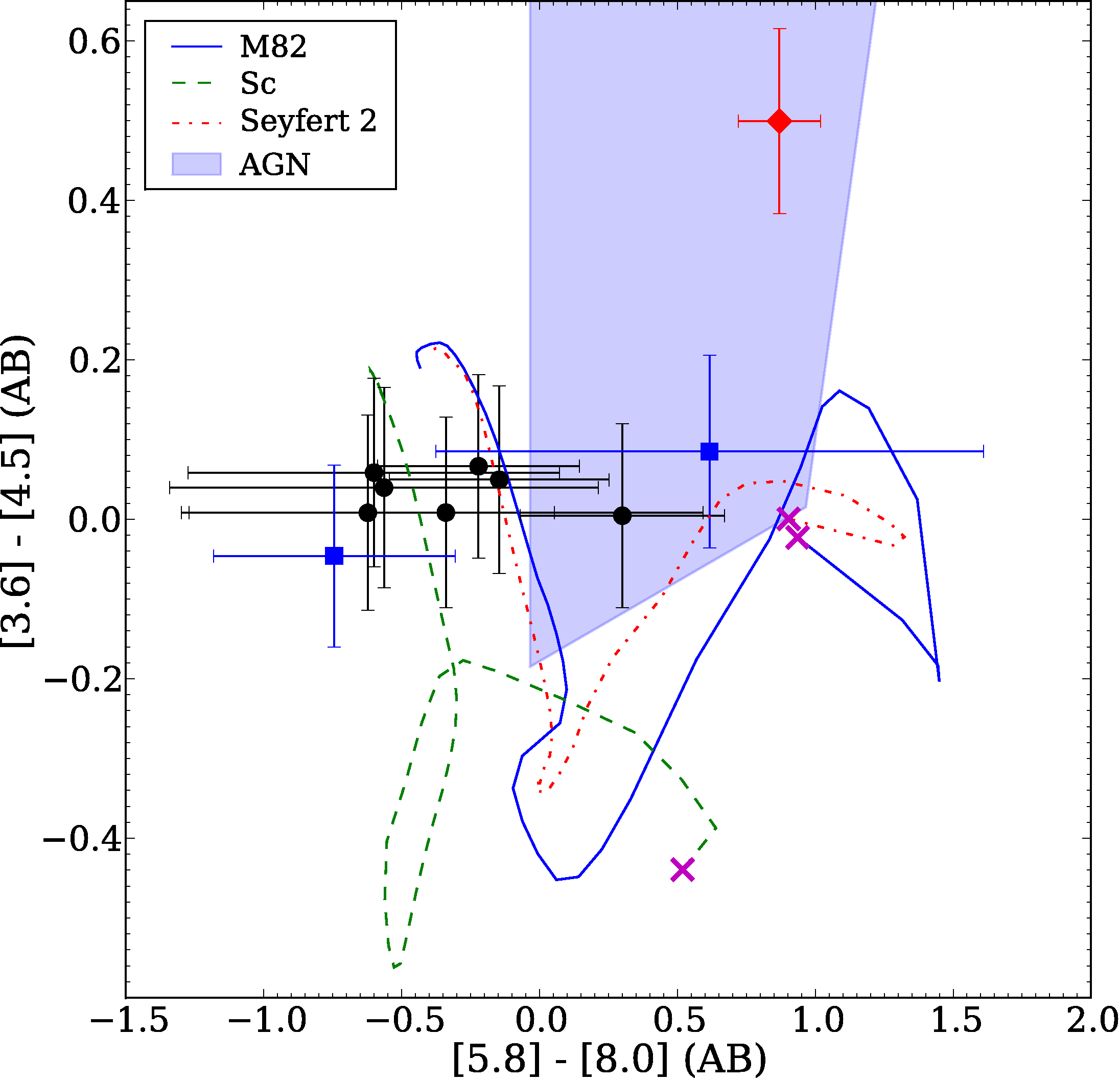}
\end{center}
\caption{IRAC color--color plot of the 24~$\micron$ emitting cluster members (circles). Overplotted are
non-evolving tracks of various spectral templates as they are redshifted from $z=0$ to $z=2$ (see legend;
the crosses indicate the $z=0$ end of each track), taken from the library of \citet{Polletta_2007}. Clearly,
the colors of most of the cluster 24~$\micron$ sources are not consistent with those expected of Type I 
QSOs (shown by the shaded area marked `AGN' in the legend), and are more similar to those expected of star
forming galaxies at this redshift. The X-ray/24~$\micron$ source (SSTU~J221600.14-173836.8; see
Section~\ref{s_sourceMatching}; marked with a diamond) occupies the same region of the color--color plane as
the QSOs. The positions of the X-ray point sources that are not 24~$\micron$ sources (PS1 and PS2; see
Section~\ref{s_XRayAnalysis}) are marked with squares.}
\label{f_IRACColCol}
\end{figure}

\subsection{Infrared Colors}
\label{s_AGN}
Re-radiation by dust is the source of the 24~$\micron$ flux from the cluster mid-infrared sources, but this
emission can be powered primarily by either star formation or AGN. Studies of both the field
\citep[e.g.][]{PerezGonzalezPaths_2008} and lower redshift clusters \citep[e.g.][]{Geach_2006, Marcillac_2007}
suggest that $<10$\% of the 24~$\micron$ emitters are likely to be powered by AGN rather than star formation;
here we perform some simple tests to determine the nature of the sources in J2215.

Luminous ($L_{\rm X (2-10 \ keV)} \gtrsim 10^{42}$ erg s$^{-1}$) X-ray point sources are
powered by AGN rather than starburst activity \citep[e.g.][]{Georgakakis_2007}. As mentioned
above (Section~\ref{s_sourceMatching}), none of the sample of 24~$\micron$ emitting cluster
members are associated with X-ray point sources detected in our \textit{Chandra} data, which reaches a
limiting flux of $\approx 1.0 \times 10^{-16}$~erg~s$^{-1}$~cm$^{-2}$ (roughly corresponding
to $L_{\rm X(2-10 \ keV)} >  0.8 \times 10^{42}$~erg~s$^{-1}$ at the cluster redshift,
assuming spectral index $\alpha = 2$ and $N_{\rm H} = 1 \times 10^{22}$~cm$^{-2}$).

\begin{deluxetable*}{cccccccc}
\tablewidth{0pt}
\tabletypesize{\scriptsize}
\tablecaption{Mid-infrared derived star formation rates for 24~$\micron$ emitting members of
J2215.9-1738\label{t_LIRSFR}}
\tablehead{\colhead{SSTU} & \colhead{Alt. ID\tablenotemark{a}} & \colhead{$f_{\rm 24}$} & \colhead{CE01
$L_{\rm IR}$} & \colhead{CE01 SFR} & \colhead{DH02 $L_{\rm IR}$} & \colhead{DH02 SFR} & \colhead{r}\\ &  &
($\mu$Jy)  & ($\times 10^{11}$ L$_\sun$) & (M$_\sun$ yr$^{-1}$) & ($\times 10^{11}$ L$_\sun$) & (M$_\sun$
yr$^{-1}$) & (Mpc)}
\startdata
\sidehead{Spectroscopic members:}
J221557.24-173753.2\phantom{0} & 35 & 159 $\pm$ 10 & \phantom{0}9.0$^{+8.0}_{-2.8\phantom{0}}$ &
160$^{+140}_{-50}$ &
\phantom{0}7.7$^{+23.8}_{-2.5}$ & 130$^{+410}_{-40}$ & 0.17\\
J221556.92-173804.7\phantom{0} & 39 & \phantom{0}98 $\pm$ 9\phantom{0} &
\phantom{0}5.6$^{+4.7}_{-1.7\phantom{0}}$ &
100$^{+80}_{-30\phantom{0}}$ & \phantom{0}4.7$^{+14.1}_{-1.5}$ & \phantom{0}80$^{+240}_{-30}$ & 0.19\\
J221600.38-173750.5\tablenotemark{*} & 53 & 159 $\pm$ 9\phantom{0} & 9.0$^{+7.9}_{-2.8}$ &
155$^{+135}_{-47}$
& \phantom{0}7.7$^{+23.4}_{-2.5}$ & 131$^{+403}_{-42}$ & 0.25\\
J221609.88-173826.4\phantom{0} & 371 & \phantom{0}93 $\pm$ 9\phantom{0} &
\phantom{0}5.3$^{+4.9}_{-1.7\phantom{0}}$ &
\phantom{0}90$^{+80}_{-30\phantom{0}}$ & \phantom{0}4.5$^{+14.3}_{-1.5}$ & \phantom{0}80$^{+250}_{-30}$ &
1.39\\
J221558.94-174101.2\phantom{0} & 400 & 122 $\pm$ 12 & \phantom{0}7.0$^{+6.7}_{-2.2\phantom{0}}$ &
120$^{+120}_{-40}$ &
\phantom{0}6.0$^{+19.5}_{-2.0}$ & 100$^{+340}_{-30}$ & 1.51\\
\sidehead{Photometric members:}
J221558.22-173822.1\phantom{0} & 899 & 108 $\pm$ 9\phantom{0} & \phantom{0}6.2$^{+5.7}_{-1.9\phantom{0}}$ &
110$^{+100}_{-30}$ & \phantom{0}5.2$^{+16.6}_{-1.7}$ & \phantom{0}90$^{+290}_{-30}$ & 0.17\\
J221556.80-173721.9\phantom{0} & 529 & 244 $\pm$ 10 & 13.9$^{+12.7}_{-4.3}$ & 240$^{+220}_{-70}$ & 
11.8$^{+37.4}_{-3.9}$ & 200$^{+640}_{-70}$ & 0.40\\
J221554.62-173855.0\phantom{0} & 1118 & \phantom{0}73 $\pm$ 9\phantom{0} &
\phantom{0}4.2$^{+3.8}_{-1.3\phantom{0}}$ &
\phantom{0}70$^{+70}_{-20\phantom{0}}$ & \phantom{0}3.5$^{+11.2}_{-1.2}$ & \phantom{0}60$^{+190}_{-20}$ &
0.65\\
\sidehead{Unconfirmed sources:}
J221559.68-173758.9\phantom{0} & 744/747 & 239 $\pm$ 9\phantom{0} & 13.7$^{+13.1}_{-4.3}$ &
240$^{+230}_{-70}$
&
11.6$^{+38.0}_{-3.8}$ & 200$^{+650}_{-70}$ & 0.15\\
\enddata
\tablecomments{Columns labelled CE01 have quantities derived using the \citet{CharyElbaz_2001} spectral
templates; likewise, quantities in columns labelled DH02 are calculated using the \citet{DaleHelou_2002}
templates. The minimum and maximum derived infrared luminosities from each template library are quoted as 
the error bars on $L_{\rm IR}$ and SFR.}
\tablenotetext{a}{The ID number in this column corresponds to the ID given in Table~\ref{t_redshifts} for
objects with spectroscopic redshifts. For objects with photometric redshifts and SSTU~J221559.68-173758.9,
this ID number corresponds to that in the catalog of \citet{Hilton_2009}.}
\tablenotetext{*}{This object is more likely to be an AGN rather than a star forming galaxy, based on its
IRAC
colors (see Section~\ref{s_AGN}).}
\end{deluxetable*}

Another way to identify AGN is through their infrared colors \citep{Stern_2005}. Figure~\ref{f_IRACColCol}
shows the [3.6]-[4.5], [5.8]-[8.0] color--color plot of the 24~$\micron$ emitting cluster members.
Unfortunately, the uncertainties on the [5.8]-[8.0] colors are large: the mean error
is $\approx 0.6$~mag, which limits our ability to draw firm conclusions from this test. This is driven by the
fact that only three of the 24~$\micron$ sources are detected at $>3\sigma$ in the 8.0$~\micron$ channel. One
source (SSTU~J221558.94-174101.2) is not detected at either 5.8~$\micron$ or 8.0~$\micron$. We
overplot in this figure the non-evolving tracks of several spectral templates taken from the library of
\citet{Polletta_2007}. The shaded area shows the region of color--color space occupied primarily by
broad-lined AGN according to the criteria of \citet{Stern_2005}. Figure~\ref{f_IRACColCol} shows
that only one of the cluster 24~$\micron$ sources falls within this region (SSTU~J221600.38-173750.5). The
other 24~$\micron$ sources have colors consistent with star forming galaxies at this redshift.

We also plot the positions of the X-ray point sources identified in the \textit{Chandra} data
(Section~\ref{s_XRayAnalysis}) in Figure~\ref{f_IRACColCol}. The source which is detected in both
the X-ray and 24~$\micron$ data is very red in [5.8]-[8.0] color, consistent with it being an obscured, 
high redshift AGN. Of the two \textit{Chandra} sources that are spectroscopically confirmed cluster members,
one lies inside the QSO region of color--color space, whereas the other has similar colors in the IRAC bands
to the cluster 24~$\micron$ sources. However, as with the 24~$\micron$ sources, the color uncertainties are
large.

The IRAC colors do not allow us to distinguish between normal galaxies and narrow lined AGN such
as the Seyfert 2 class, as shown by the template tracks in Figure~\ref{f_IRACColCol}. However, galaxies of
this class are often host to significant star forming activity, generally being dominated by signatures of
star formation at optical and infrared wavelengths \citep[e.g.][]{DellaCeca_2001, CidFernandes_2004,
Panessa_2005}.

\subsection{Star Formation}
\label{s_SFR}
We now proceed to estimate infrared luminosity ($L_{\rm IR}$) derived star formation rates for cluster
24~$\micron$ sources that were not classified as AGN due to either their infrared colors or the presence of
X-ray emission. We include the `unconfirmed source' (SSTU~J221559.68-173758.9,
Section~\ref{s_sourceMatching} above and Figure~\ref{f_confused}) and the source undetected in the
5.8~$\micron$ and 8.0~$\micron$ channels (SSTU~J221558.94-174101.2) in this sample. We first derive
8-1000~$\micron$ luminosities ($L_{\rm IR}$) from the measured 24~$\micron$ flux densities using the full
range of templates in the libraries of \citet{CharyElbaz_2001} and \citet{DaleHelou_2002}. Throughout this
analysis we treat each template library independently. Because we lack the longer wavelength data needed to
constrain the SEDs, we take the median $L_{\rm IR}$ value derived from each template set as the best estimate
of $L_{\rm IR}$, and adopt the minimum and maximum values of $L_{\rm IR}$ as error bars. These
systematic errors dominate over the random error due to uncertainty in the 24~$\micron$ flux measurements. The
$L_{\rm IR}$ values are then converted into star formation rates using the \citet{Kennicutt_1998} law. The
derived infrared luminosities and star formation rates for each cluster galaxy in the sample are listed in
Table~\ref{t_LIRSFR}.

As Table~\ref{t_LIRSFR} shows, all of the 24~$\micron$ sources in J2215 can be classified as Luminous Infrared
Galaxies (LIRGs; galaxies with $L_{\rm IR} > 10^{11}$ $L_\sun$), and have very high star formation rates,
typically $\sim 100$ M$_\sun$ yr$^{-1}$. If the unconfirmed 24~$\micron$
source (Section~\ref{s_sourceMatching} above) is actually a cluster member, it is
an Ultra Luminous Infrared Galaxy (ULIRG; $L_{\rm IR} > 10^{12}$ $L_\sun$) -- although this luminosity may be
overestimated somewhat due to the presence of a neighboring galaxy within $< 1\arcsec$
(Figure~\ref{f_confused}). Neglecting the unconfirmed 24~$\micron$ source, there are three objects
at projected clustercentric radii $< 250$ kpc with very high SFRs, and so intense star formation may be
taking place in the heart of the cluster, as also hinted at by the narrowband \textsc{[O ii]} imaging of
\citet{Hayashi_2010}. Studies of the environmental dependence of star formation rate in the field have shown
that SFR actually increases with increasing local galaxy density at $z\sim 1$, whereas the opposite trend is
observed in the local universe \citep{Elbaz_2007, Cooper_2008}. However, these studies also show that the SFR
falls off beyond a critical peak galaxy density. There is $\approx 1.5$~Gyr difference in look back time
between the redshift of J2215 and $z=1$, and so perhaps it is possible that we are seeing the peak of the
local density--SFR relation shift to higher densities as redshift increases.

The median values of $L_{\rm IR}$ and SFR that we estimate for each galaxy are consistent between both the
template libraries of \citet{CharyElbaz_2001} and \citet{DaleHelou_2002}, although we note that the
upper limit of $L_{\rm IR}$ and SFR estimated using the \citet{DaleHelou_2002} templates is significantly
higher. It is possible that the lower limit on the values in Table~\ref{t_LIRSFR} may be overestimated, if
neither of the template libraries, which are based on observations at considerably lower redshift, are
representative of galaxies at $z\approx 1.5$. \citet{Murphy_2009} show that this is likely to be the case from
mid-infrared spectroscopic observations of galaxies in this redshift regime. In particular, the equivalent
widths of features due to polycyclic aromatic hydrocarbons (PAHs) seem to be underestimated by the
\citet{CharyElbaz_2001} templates compared to observed high redshift galaxies, and for galaxies with $L_{\rm
IR} > 3 \times 10^{13}$ M$_\Sun$ and $z > 1.4$ this leads to an overestimate of $L_{\rm IR}$ by a factor $\sim
5$ when $L_{\rm IR}$ is derived from 24~$\micron$ data alone. For J2215 at $z=1.46$, a broad PAH feature at
8.6~$\micron$ is redshifted into the MIPS 24~$\micron$ band, and so this may be an issue for the measurements
reported in Table~\ref{t_LIRSFR}. Longer wavelength data (e.g. at 70~$\micron$) are required to investigate
this issue further, and it has been shown that such data significantly improves the constraints on $L_{\rm
IR}$ inferred from template fitting, particularly at high redshifts (see Figure~2 of
\citealt{Murphy_2009}).

Other lower redshift studies have used 24~$\micron$ observations to quantify the fraction of star forming
galaxies $f_{\rm SF}$ in clusters, finding $f_{\rm SF}$ to increase from $\approx 3$\% at $z\approx 0$ to
$\approx 13$\% by $z\approx 0.8$ \citep[e.g.][]{Saintonge_2008, Haines_2009}. Given the depth of our
24~$\micron$ observations, and the fact that sample definitions used to estimate $f_{\rm SF}$ vary, we are
only able to attempt a very rough estimate of $f_{\rm SF}$ in J2215. We find $f_{\rm SF} \sim 16$\% for
galaxies within $R_{200}$ and brighter than $\approx M^*+1.5$ in the galaxy luminosity function, using only
galaxies with $L_{\rm IR} > 3.3 \times 10^{11}$~$L_{\sun}$ (i.e., SFR $\gtrsim 60$~M$_\sun$~yr$^{-1}$) to
define the star forming sample. Note that we exclude the `unconfirmed source' (SSTU~J221559.68-173758.9)
in making this estimate. The higher cut in $L_{\rm IR}$ used here compared to that used in e.g.
\citet{Haines_2009} to define $f_{\rm SF}$ ($L_{\rm IR} > 5 \times 10^{10}$~$L_{\sun}$) suggests that the star
forming fraction in J2215 may be even higher, if an equivalent, lower cut in $L_{\rm IR}$ was used.

\subsection{Optical Properties}
\begin{figure}
\begin{center}
\includegraphics[width=8.5cm]{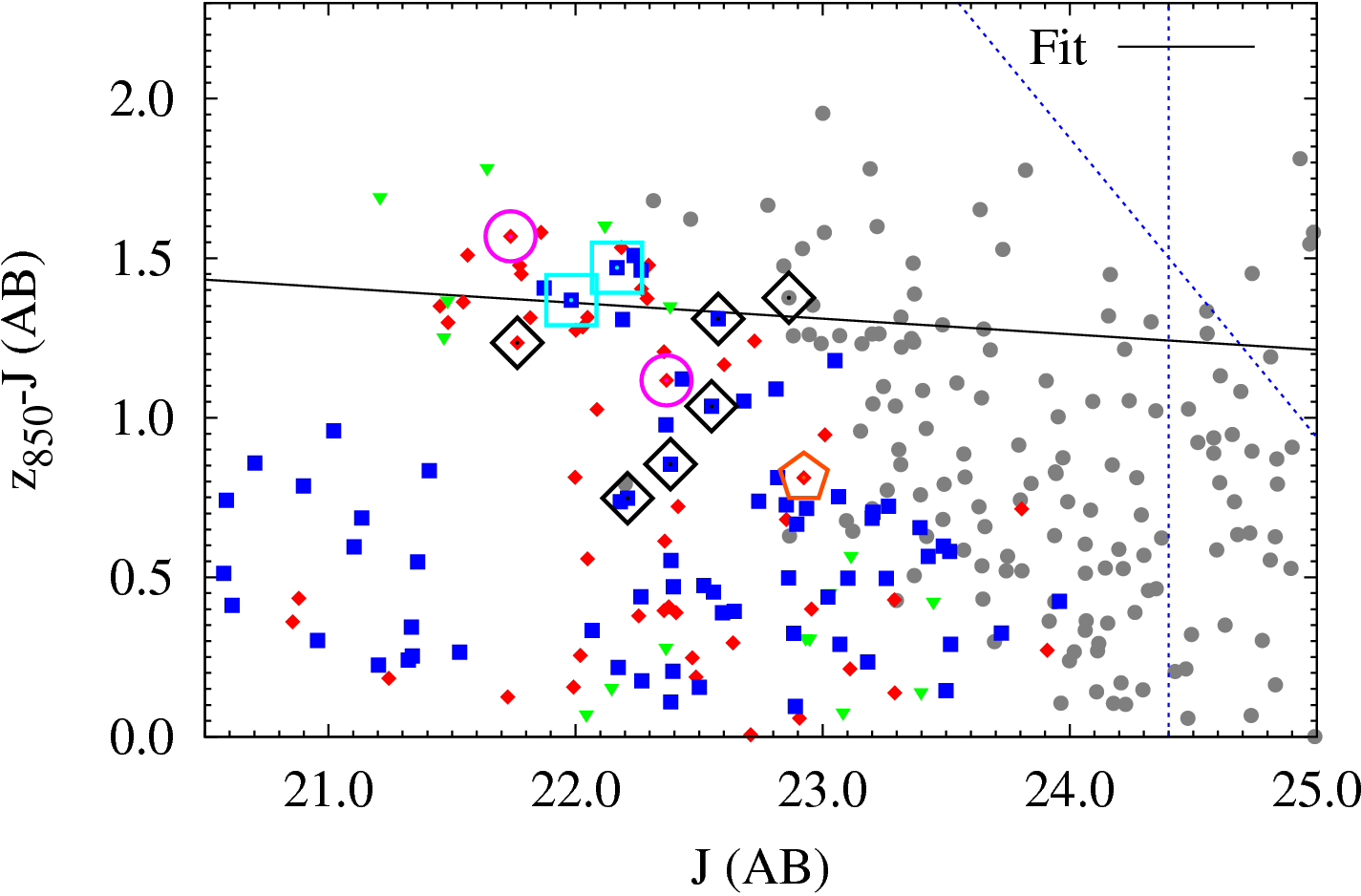}
\end{center}
\caption{$z_{850}-J$ color--magnitude diagram of all galaxies within 0.8~Mpc of the center of J2215.9-1738.
Elliptical galaxies are marked as red diamonds; S0s with green triangles; late-type galaxies with blue
squares; and galaxies for which morphologies were not determined with gray circles \citep[see][for details 
of the morphological classifications]{Hilton_2009}. The 24~$\micron$ emitting cluster members
are outlined by the large black diamonds; both possible optical counterparts to the unconfirmed 24~$\micron$
source (shown in Figure~\ref{f_confused}) by cyan squares; the X-ray emitting cluster
members by the magenta circles; and the 24~$\micron$ and X-ray source (likely to be a high redshift 
obscured AGN) by the orange pentagon. The solid black line marks the fit to the color--magnitude
relation of early type galaxies within the cluster measured by \citet{Hilton_2009}. Several of the
24~$\micron$ sources lie close to the cluster red-sequence, indicating that not all of these galaxies are
`red-and-dead'.}
\label{f_mipsCMD}
\end{figure}

\begin{figure}
\begin{center}
\includegraphics[width=8cm]{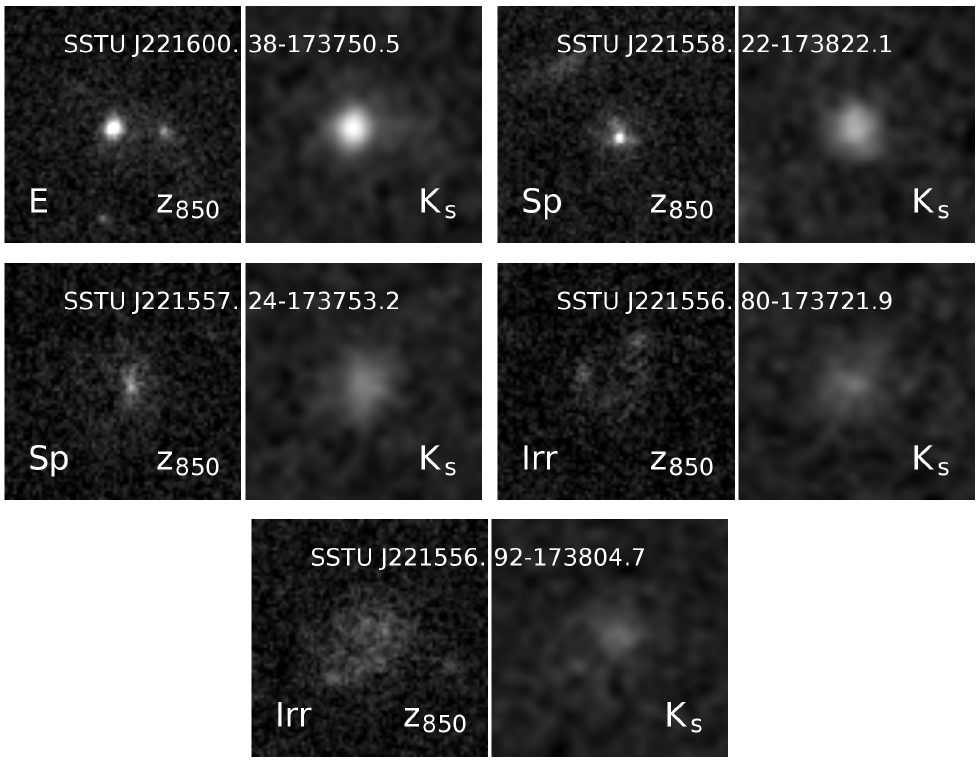}
\end{center}
\caption{ACS $z_{850}$ and MOIRCS $K_s$ postage stamp images of 24~$\micron$ emitting members of 
J2215.9-1738 that fall within the ACS/MOIRCS imaging area \citep[see][for details]{Hilton_2009}. Each image 
is $3.75\arcsec$ on a side. Most of the galaxies are of irregular (Irr) or spiral (Sp) type. Only one of 
the galaxies is elliptical (and this object is likely to be powered by an AGN based on its infrared colors,
see Section~\ref{s_AGN}), and one galaxy fainter than $z_{850} > 24$ was not morphologically 
classified.}
\label{f_morphPlot}
\end{figure}

Using the catalog from \citetalias{Hilton_2009}, we can investigate the optical properties of the six 
24~$\micron$ emitting cluster galaxies within $R_{200}$, in addition to the objects detected in the
\textit{Chandra} X-ray observations. Figure~\ref{f_mipsCMD} shows the $z_{850} - J$
color--magnitude diagram of the cluster with the 24~$\micron$ cluster members highlighted. Three of these
galaxies are within $< 2.3\sigma$ of the cluster red-sequence. Lower redshift studies have also
found a significant number of 24~$\micron$ emitting, presumably star forming, cluster members close to the
red-sequence, and that the contribution of these galaxies to the total star forming population in clusters
increases with redshift \citep{Saintonge_2008}. However, the brightest of these objects in J2215 is
likely to be an AGN based on its infrared colors (SSTU~J221600.38-173750.5, see Section~\ref{s_AGN}
above). This object was also morphologically classified as an elliptical galaxy by \citetalias{Hilton_2009}.
All of the other 24~$\micron$ sources, which are most likely powered predominantly by star formation, have
late-type morphologies (spiral or irregular; see Figure~\ref{f_morphPlot}).

The two possible optical counterparts to the unconfirmed 24~$\micron$ source
(SSTU~J221559.68-173758.9, Figure~\ref{f_confused}) are highlighted in Figure~\ref{f_mipsCMD}. These objects
are both slightly redder than the cluster red-sequence, but lie within $<1\sigma$ of it, and were
morphologically classified as late type galaxies by \citetalias{Hilton_2009}. If this source is truly within
the cluster, it is therefore likely to be an example of a red-sequence galaxy with high SFR.

We also plot the positions of the X-ray sources in Figure~\ref{f_mipsCMD}. The cluster member X-ray sources
(PS1 and PS2; Section~\ref{s_XRayAnalysis}) are both within $<2\sigma$ of the red-sequence, whereas the
high redshift, obscured QSO in the background of the cluster is considerably bluer than the red-sequence.

\subsection{Stellar Population Modelling and Specific Star Formation Rates}
We investigated the stellar populations of the 24~$\micron$ sources located within $R_{200}$ by
fitting their broadband spectral energy distributions (SEDs) to the models of \citet{BruzualCharlot_2003}. We
combined the IRAC photometry (Section~\ref{s_IRAC} above) with the rest-frame optical photometry
presented in \citetalias{Hilton_2009}, using the \texttt{SExtractor} \texttt{MAG\_AUTO} magnitudes in the
latter case as estimates of total magnitude. Note that the cross matching between the IRAC and the $K_s$
selected catalog described in Section~\ref{s_IRAC} showed that each of the six 24~$\micron$ sources is
uniquely associated with a single corresponding object across both catalogs. However, the IRAC photometry
for one object (SSTU~J221558.22-173822.1) suffers from blending with another galaxy $\approx 2.5\arcsec$
away. The blended galaxy is $\approx1$~mag fainter at $K_s$, and so we expect that this should lead to only a
small overestimate of the stellar mass of this object.

We followed a similar approach to \citet{Shapley_2005} in performing the SED fitting. We use a grid of solar
metallicity models with exponentially declining star formation histories with 20 values of $\tau$ in the range
0.1--20 Gyr, and 50 ages in range 0.001--4 Gyr, i.e. constrained such that the maximum allowed age is less
than that of the universe at $z=1.46$. We adopt a \citet{Chabrier_2003} Initial Mass Function (IMF). We
include the effect of dust extinction using the \citet{Calzetti_2000} law, allowing the value of $E(B-V)$ to
vary between 0.0--0.5 in steps of 0.02. We fitted the SEDs by analytically calculating the normalization $N$
for each model SED where $d\chi^2/dN=0$, adopting the model with the lowest $\chi^2$ value as the best fit.
The stellar mass $M_*$ is then estimated from the value of $N$. We therefore fit for a total of four
parameters (age, $\tau$, $E(B-V)$, $M_*$).

We estimated the errors on parameters through Monte-Carlo simulations; we generated 1000 fake SEDs for each
object by replacing each flux measurement with a random variate drawn from a Gaussian distribution consistent
with the observed photometric uncertainties, and fitted them using the same method used for the real data as
described above. We adopt 68.3-\%-ile confidence intervals as our error estimates, measured as the
corresponding fraction of the fits to the Monte-Carlo realizations.

\begin{figure}
\begin{center}
\includegraphics[width=8.5cm]{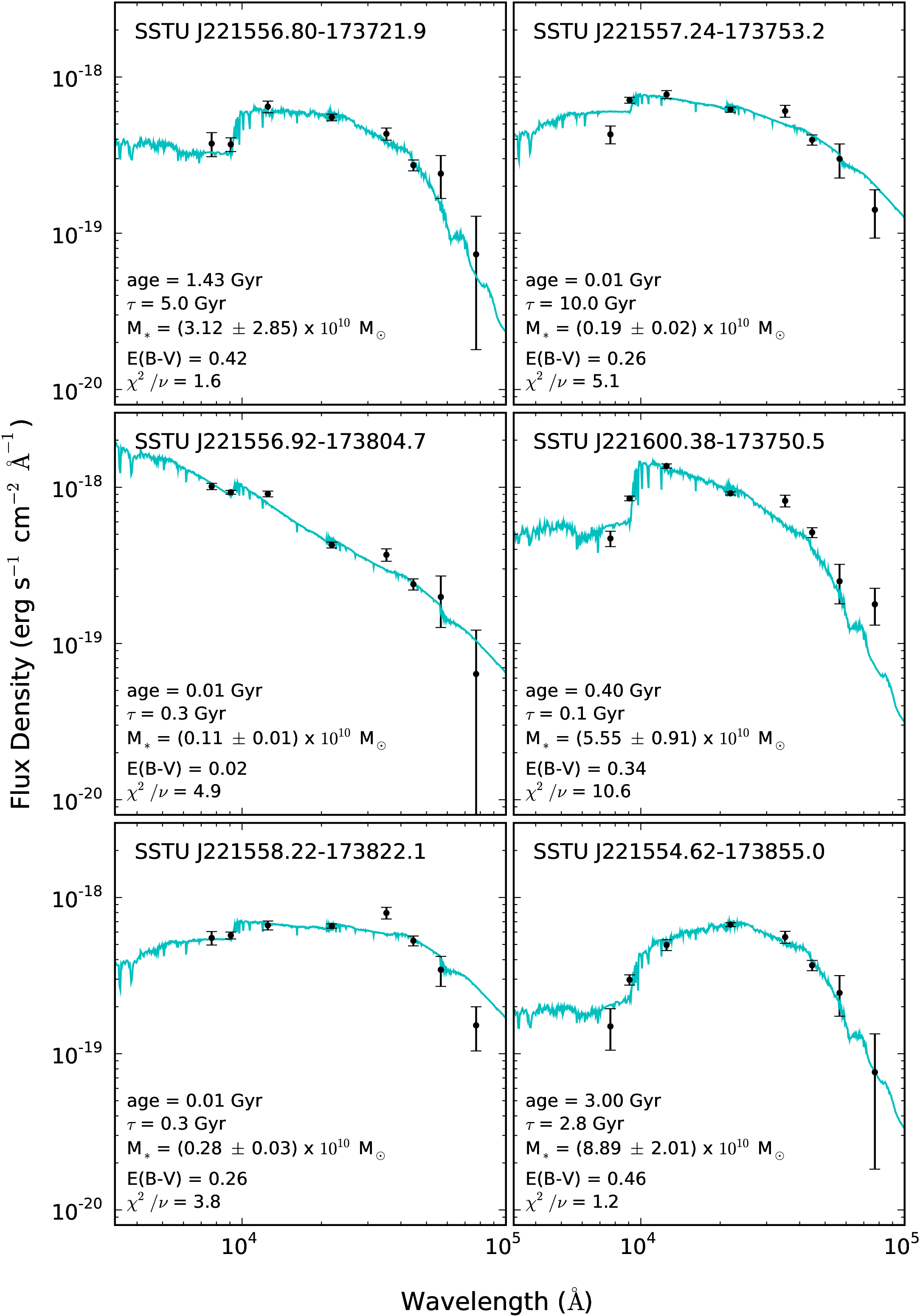}
\end{center}
\caption{Broadband spectral energy distributions of the six 24~$\micron$ emitting cluster members with
available rest-frame optical photometry from \citet{Hilton_2009} and IRAC photometry from this work 
(points). The best fitting \citet{BruzualCharlot_2003} model is shown by the solid line. Note that the 
object with the highest $\chi^2/\nu$ value (J221600.38-173750.5) is likely to host an AGN based on its
infrared colors (Section~\ref{s_AGN}).}
\label{f_SEDsPlot}
\end{figure}

Figure~\ref{f_SEDsPlot} shows the results. The SEDs of the 24~$\micron$ emitting cluster members are
in most cases best fit by young (age $< 400$ Myr) stellar populations, with large dust extinction values
($E(B-V) > 0.2$). However, there are many degeneracies between the fitted parameters \citep[such as between
$\tau$ and $E(B-V)$; see the discussion in e.g.][]{Shapley_2005} and different stellar population models yield
systematically different results; for example, if we instead fit the SEDs using the \citet{Maraston_2005}
models, we generally obtain higher $E(B-V)$ values and younger ages. We note that in most cases the reduced
$\chi^2$ values indicate that the fits are not especially good. The worst case (J221600.38-173750.5 with
$\chi^2/\nu = 10.6$) was classified as an AGN based on its infrared colors (see Section~\ref{s_AGN}
above), and so the poor fit for this object is not surprising.

\begin{figure}
\begin{center}
\includegraphics[width=8cm]{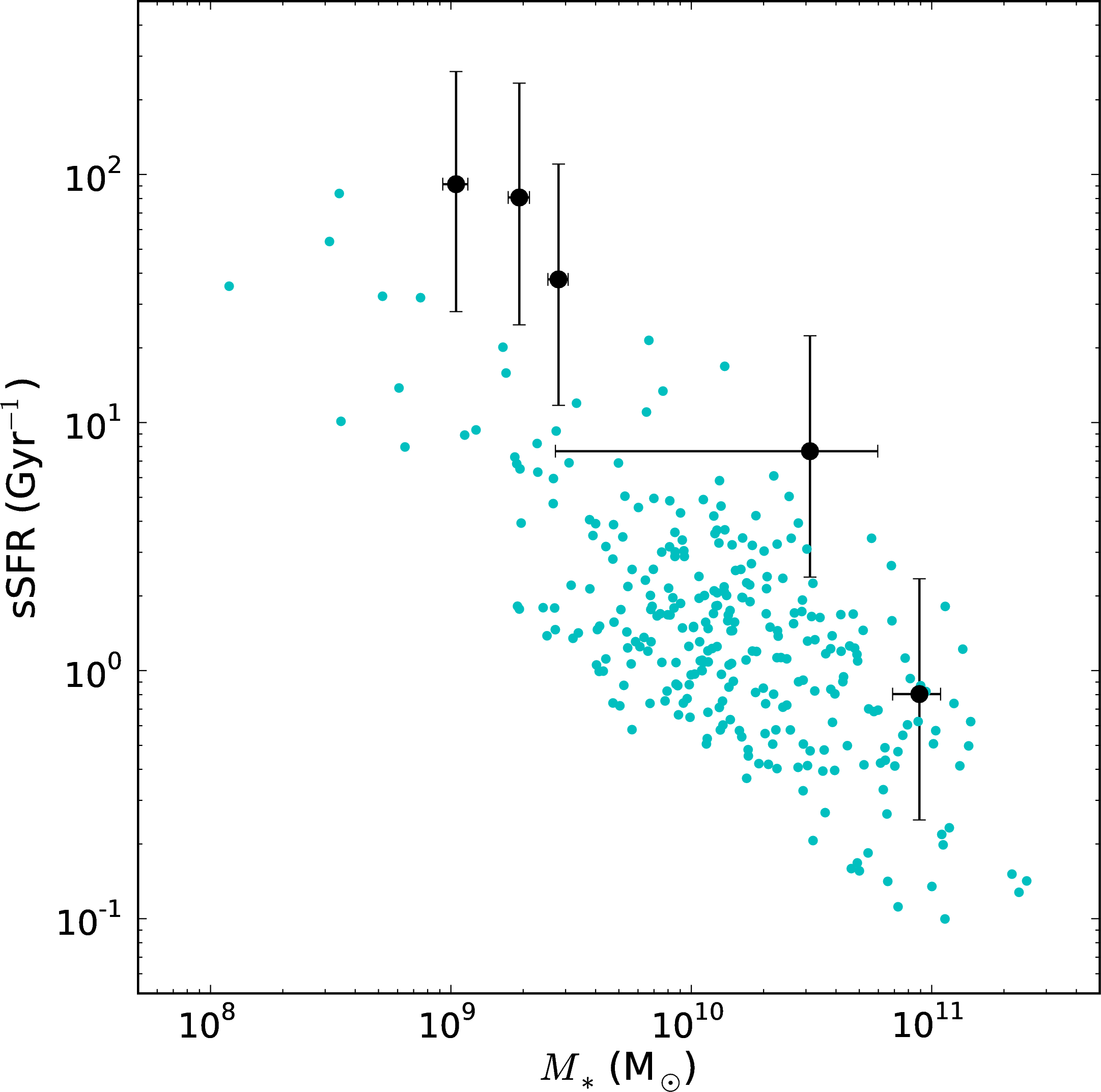}
\end{center}
\caption{Specific star formation rate (sSFR) versus stellar mass ($M_*$) for the five 24~$\micron$
emitting cluster members not classified as AGN plotted in Figure~\ref{f_SEDsPlot} (large points with error
bars). The data from the $1.0 <z < 1.5$ field sample of \citet{Santini_2009} are also plotted for comparison
(small points). Note that the values for the field sample have been scaled to a \citet{Chabrier_2003} IMF, 
as was used in deriving the values for the cluster galaxies. The 24~$\micron$ emitting cluster galaxies 
appear to follow a similar sSFR--$M_*$ relation to field galaxies at this redshift, though the error bars 
are large. The SFRs for the cluster galaxies, estimated from the 24~$\micron$ fluxes alone, may be
overestimated (see the discussion in Section~\ref{s_SFR}).}
\label{f_sSFR}
\end{figure}

A number of authors have found that SED-based stellar mass estimates are reasonably robust, despite
degeneracies between other stellar population parameters \citep[e.g.][]{Shapley_2005, Muzzin_2009}. We used
the stellar mass measurements derived from the SED fitting to estimate specific star formation rates
(sSFR) for the cluster 24~$\micron$ sources. Figure~\ref{f_sSFR} shows the results. Despite the very small
sample size, we recover the well known result that the lowest mass galaxies are building up their stellar
mass most rapidly, having the highest sSFRs. The cluster 24~$\micron$ sources follow a similar
sSFR--$M_*$ relation to that measured in the field at $1.0 < z < 1.5$ \citep[e.g.][]{Santini_2009}, although
the sSFRs of the cluster sources appear to be biased high compared to the field sample. However, the
field galaxy sample contains relatively fewer objects at $z\sim 1.5$ compared to the low end of the sample
redshift range, and the locus of the relation moves upward in Figure~\ref{f_sSFR} as redshift increases. It is
also possible that the cluster galaxy SFRs, which are derived from the 24~$\micron$ fluxes alone, may be
overestimated (see the discussion in Section~\ref{s_SFR}).

\section{Discussion}
\label{s_discussion}

\subsection{The Impact of Unresolved Point Sources On High-Redshift X-ray Cluster Surveys}
As the results of this paper show, the presence of emission from unresolved point sources in
X-ray observations of high-redshift clusters of galaxies can lead to significant biases in temperature
measurements if unaccounted for. As cosmological analyses of X-ray cluster surveys rely on temperature as a
mass proxy, the increase in the X-ray AGN population with redshift is therefore likely to also bias any
cosmological interpretation that includes such high redshift objects. This issue is not likely to affect XCS,
as it is planned to exclude $z > 1$ clusters from the cosmological analysis, as few such systems will be
detected \citep{Sahlen_2009}. However, this may be an issue for larger scale, low resolution X-ray surveys
such as the upcoming all-sky \textit{eROSITA} mission \citep{Predehl_2010}, which will detect tens of
thousands of galaxy clusters out to $z \approx 1.3$, including $\sim 500$ at $z > 1$. To extract the full
cosmological information from this sample will therefore require either a follow-up program of high resolution
pointed observations of detected high redshift clusters, or could alternatively perhaps be addressed
statistically, using the existing data in the \textit{Chandra} archive to derive the amount of
contamination by AGN in clusters as a function of redshift \citep[e.g.][]{Martini_2009, Gilmour_2009}.

\subsection{Star Formation in High Redshift Galaxy Clusters}
The results of our analysis suggest that the core of J2215 may contain galaxies with very high SFRs,
as 24~$\micron$ sources which are likely to be powered by star formation are found within projected distances
of $<250$~kpc of the cluster X-ray position. Significant star forming activity in the cluster core is also
suggested by the narrowband \textsc{[O ii]} observations by \citet{Hayashi_2010}, which are sensitive to
galaxies with lower SFRs ($>2.6$~M$_\sun$~yr$^{-1}$) than is probed by our MIPS observations.

Studies of clusters at lower redshift have revealed a significant amount of obscured star formation in
clusters; however, this activity seems to take place mostly in the infall regions, and not in the cluster
cores \citep{Geach_2006, Marcillac_2007, Saintonge_2008, Haines_2009, Koyama_2010}. This suggests that
triggered star formation, driven by the interaction of infalling galaxies with the cluster potential, is one
of the main channels for transforming star forming, late type galaxies in clusters into passively evolving
early type galaxies. In this scenario, the gas within infalling galaxies is compressed upon entering the
cluster, leading to a burst of star formation which rapidly uses up the available gas reservoir, eventually
leading to star formation being quenched. If this is the general case, then why are high SFRs observed in the
center of J2215?

Perhaps environmental differences between clusters are responsible, as the amount of star formation in
clusters is seen to vary significantly from cluster to cluster \citep[e.g.][]{Geach_2006}. At low redshift,
there are hints that the dynamical state of the cluster plays a role. \citet{Braglia_2009} examined star
formation in two $z\approx 0.3$ clusters of opposite dynamical state: one of their targets was a relaxed,
cool core cluster, while the other appears to be a merger. They found a significantly increased fraction of
star forming galaxies within $R_{200}$ in this latter system. Similarly, \citet{HainesAbell_2009} examined
star formation in the merging system A1758 at $z=0.28$, finding significant dust obscured star formation
taking place within $<500$~kpc of the core of one of the components of this system, suggesting that cluster
mergers can trigger star formation in cluster galaxies.

Although J2215 is not resolved into two kinematically distinct components with the current spectroscopic
sample of members (Section~\ref{s_velDist}, above), the limited data currently available hints that the
cluster may be dynamically somewhat disturbed. In addition to the cluster velocity distribution, the cluster
lacks an obvious BCG, and the BCG candidate selected in the study by \citet{Collins_2009} is located at a
significant offset of $\approx 300$~kpc from the cluster X-ray position. 

It will be interesting to compare our results for J2215 to the properties of star forming galaxies in the few
other $z > 1.3$ clusters that are currently known \citep{Mullis_2005, Stanford_2005, Eisenhardt_2008,
Wilson_2009, Papovich_2010, Tanaka_2010}. In particular, XMMU J2235.3-2557 at $z=1.39$ \citep{Mullis_2005}
appears to be a very relaxed and mature system already at this epoch, showing evidence of a cool core X-ray
morphology \citep{Rosati_2009}, and the core of this cluster is dominated by a group of massive, passively
evolving galaxies, the brightest of which are significantly brighter than the BCG in J2215 \citep{Lidman_2008,
Hilton_2009}. Furthermore, this cluster appears to be $\approx 3$ times more massive than J2215 based on a
comparison of the cluster X-ray temperatures \citep{Rosati_2009, Jee_2009}, and may therefore be expected to
have a lower fraction of star forming galaxies than J2215 if star formation is truncated earlier in more
massive systems.

\section{Conclusions}
\label{s_conclusions}
We have explored the AGN and star forming populations of the cluster J2215.9-1738 at $z=1.46$ using high
resolution X-ray data from the \textit{Chandra} satellite and infrared observations from the \textit{Spitzer
Space Telescope}. This is the first study of star formation as traced by mid-infrared observations in
a cluster at $z \approx 1.5$. We found:

\begin{enumerate}

\item{The cluster emission is contaminated by X-ray point sources, leading to the X-ray temperature being
overestimated in the analysis presented in \citetalias{Stanford_2006}; however, these point sources only 
contribute $\approx 15$\% to the total flux. Two of the X-ray point sources revealed by the \textit{Chandra}
observations are cluster members, while a third has properties consistent with it being a high redshift,
highly obscured AGN in the background. The cluster has temperature $T=4.1_{-0.9}^{+0.6}$~keV and
bolometric luminosity $L_{\rm X} = (2.92_{-0.35}^{+0.24}) \times 10^{44}$~erg~s$^{-1}$ (extrapolated to 2~
Mpc radius) from a joint analysis of the \textit{XMM-Newton} and \textit{Chandra} data. The cluster is less
luminous than expected from self similar evolution of the local $L_{\rm X}-T$ relation at the $\approx
2\sigma$ level.}

\item{The velocity dispersion of the cluster is measured to be $\sigma_v = 720 \pm 110$~km~s$^{-1}$, from
31 galaxies within $R_{200}$. There is no clear evidence that the cluster velocity distribution is composed of
two kinematically distinct components, although the \citet{Hartigan2_1985} dip test of unimodality still
hints that the cluster is not completely relaxed. Following the revised X-ray analysis presented in this
paper, the cluster is found to lie on the $\sigma_v-T$ relation, contrary to the result reported in
\citetalias{Hilton_2007}.}

\item{Mid-infrared imaging reveals a total of eight $>5 \sigma$ 24~$\micron$ sources that are cluster members,
selected using spectroscopic or photometric redshifts. In addition, there is a prominent 24~$\micron$ source
(J221559.68-173758.9) with two possible optical/near-infrared counterparts located $\approx 17\arcsec$ from
the cluster center that may be associated with the cluster. One of the 24~$\micron$ sources is found
to have the infrared colors expected of an AGN, and has an elliptical morphology. The remaining objects are
most likely powered by star formation, and if this is the case have SFRs $\sim 100$~M$_\sun$~yr$^{-1}$,
adopting the \citet{CharyElbaz_2001} spectral templates and assuming that the \citet{Kennicutt_1998} law holds
at this redshift.}

\item{The cluster member AGN identified in the X-ray and infrared observations are located within $\lesssim 2
\sigma$ of the cluster red-sequence, as are three (possibly four) 24~$\micron$ sources assumed to be
powered by star formation. Three of the 24~$\micron$ sources are also found within a projected distance of $<
250$~kpc from the cluster center, suggesting that the core of J2215 may be host to galaxies with very high
star formation rates, in contrast to clusters at lower redshift.}

\end{enumerate}

\acknowledgments

We thank the referee for a number of suggestions that improved the clarity of this paper. We thank Paola
Santini for providing the field galaxy data plotted in Figure~\ref{f_sSFR}. This work is based
in part on observations with the \textit{Spitzer Space Telescope}, which is operated by the Jet Propulsion
Laboratory, California Institute of Technology, under NASA contract 1407, and on observations with the
\textit{Chandra} X-ray Observatory. The W. M. Keck Observatory is a scientific partnership between the
University of California and the California Institute of Technology, made possible by a generous gift of the
W. M. Keck Foundation. Based in part on observations obtained at the Gemini Observatory, which is operated by
the Association of Universities for Research in Astronomy, Inc., under a cooperative agreement with the NSF on
behalf of the Gemini partnership: the National Science Foundation (United States), the Science and Technology
Facilities Council (United Kingdom), the National Research Council (Canada), CONICYT (Chile), the Australian
Research Council (Australia), Ministério da Ciência e Tecnologia (Brazil) and Ministerio de Ciencia,
Tecnología e Innovación Productiva  (Argentina). The analysis pipeline used to reduce the DEIMOS data was
developed at UC Berkeley with support from NSF grant AST-0071048. MH acknowledges the support of a
postdoctoral research fellowship from the University of KwaZulu-Natal. NM, AKR and ELD acknowledge financial
support from STFC. MS acknowledges financial support from the Swedish Research Council (VR) through the Oskar
Klein Centre. PTPV acknowledges the support of POCI2010 through the project POCTI/CTE-AST/58888/2004. This
work was performed under the auspices of the U.S. Department of Energy, National Nuclear Security
Administration by the University of California, Lawrence Livermore National Laboratory under contract No.
W-7405-Eng-48. The authors wish to recognize and acknowledge the very significant cultural role and reverence
that the summit of Mauna Kea has always had within the indigenous Hawaiian community; we are fortunate to have
the opportunity to conduct observations from this mountain.

\end{document}